\documentclass[preprint,12pt,3p]{elsarticle}
\usepackage{amssymb}
\usepackage{subfigure}
\usepackage{amsmath}
\usepackage{float}
\usepackage{tikz}
\usetikzlibrary{shapes.geometric}
\usetikzlibrary{shapes.arrows}
\usepackage[binary-units=true]{siunitx} 
\graphicspath {{figures/}}
\journal{}
\begin{document}

\begin{frontmatter}

\title{Prolonged Clogs in Bottleneck Simulations for Pedestrian Dynamics}

\author[label1]{Qiancheng Xu\corref{cor1}}
\address[label1]{Institute for Advanced Simulation,\\ Forschungszentrum J\"ulich, 52425 J\"ulich, Germany}

\cortext[cor1]{corresponding author}

\ead{q.xu@fz-juelich.de}

\author[label1]{Mohcine Chraibi}
\ead{m.chraibi@fz-juelich.de}

\author[label1,label2]{Armin Seyfried}
\address[label2]{School of Architecture and Civil Engineering,\\University of Wuppertal, 42119 Wuppertal, Germany}
\ead{a.seyfried@fz-juelich.de}

\begin{abstract}
This article studies clogging phenomena using a velocity-based model for pedestrian dynamics.
First, a method to identify prolonged clogs in simulations was introduced.  
Then bottleneck simulations were implemented with different initial and boundary conditions.
The number of prolonged clogs were analyzed to investigate the decisive factors causing this phenomenon.
Moreover, the time lapse between two consecutive agents passing the exit, and the trajectories of agents were analyzed.
The influence of three type of factors was studied: parameters of the spatial boundaries, algorithmic factors related to implementation of the model, and the movement model.
Parameters of the spatial boundaries include the width and position of the bottleneck exit.
Algorithmic factors are the update methods and the size of the time step.
Model parameters cover several parameters describing the level of motivation, the strength and range of impact among agents, and the shape of agents. 
The results show that the occurrence of prolonged clogs is closely linked to parameters of the spatial boundaries and the movement model but has virtually no correlation with algorithmic factors.
\end{abstract}


\end{frontmatter}


\section{Introduction}
Clogging is a phenomenon that usually arises when particles pass through narrow bottleneck structures~\cite{zuriguel2014invited}.
It is often expressed as the jamming arch formed by several interactive particles in front of the bottleneck, which significantly decreases or even stops the flow through the bottleneck.

The phenomenon occurs in different systems of inert particles such as granular material in the silo~\cite{zuriguel2014invited,arevalo2016clogging,hidalgo2013force}, dense suspension of colloidal particles~\cite{vanhecke2009jamming, zuriguel2014clogging, guariguata2012jamming,genoves2011Crystal} or electrons on the surface of liquid helium~\cite{rees2011point,rees2019commen}.
This type of clogging is usually stable if there is no external disturbance to break the balance between the particles that form the clogging~\cite{zuriguel2014invited, lazano2012break}.
Clogging can also be observed in the movement of animals~\cite{garcimartin2015flow} and humans~\cite{garcimartin2016flow} when congestion and high motivation coincide, for instance, when a large number of passengers at a train station enter carriages through a narrow train door with high motivation, or when fans at entrances to a concert hall are all trying to get in and find places near the stage~\cite{adrian2020crowds}.

Unlike with the clogging of inert particles, clogging in systems with humans is temporary.
The duration of clogs depends on the motivation level of the pedestrians involved in the clogging~\cite{zuriguel2014clogging, garcimartin2015flow, garcimartin2016flow, adrian2020crowds}.
Although the clogging of humans may last a relatively long time in some extreme cases and sometimes even leads to severe injuries~\cite{taylor1990the, krausz2012love}, in most normal cases, its duration is short even in competitive situations~\cite{adrian2020crowds}. 
In the literature, the short-term nature of the clogs is often attributed to the fluctuation in the load to the humans in the clog. 
This fluctuation, in turn, may be the result of the flexibility and elasticity of the human body.
Moreover, some clogs are avoided before forming, through complex steering mechanisms that include cognitive processes and control of the body.

However, microscopic models based on physical principles merely focus on guaranteeing volume exclusion. 
They do not take the above-mentioned factors sufficiently into account, which could lead to prolonged clogs (clogs interrupting flow for a long time) or even stable clogs similar to inert particles.
One study, \cite{kirchner2003friction}, examined this phenomenon using a cellular automaton (CA) model.
A friction parameter was introduced for an improved description of the clogging of pedestrians.
In another study, \cite{yanagisawa2009intro}, the friction parameter was extended to a function of the number of agents in clogging for a more realistic result of the pedestrian outflow through the exit.
The effect of queuing and pushing behavior in front of the bottleneck on the overall dynamics of the crowd is explored using another CA model in \cite{fischer2020micro}, where a local pushing mechanism is used \cite{yates2015incorporating}.
Another global pushing mechanism is proposed in \cite{almet2015push}.
Furthermore, game theory is combined with CA models in some studies to better reproduce the movement of pedestrians~\cite{von2015spatial, zheng2011conflict}.
Prolonged clogs and stable clogs can also be observed in the social force models for pedestrian dynamics by increasing the desired velocity of the agents~\cite{helbing2000simula}.
Introducing random behavioral variations is important to mitigate these clogs in simulations~\cite{helbing2005self, hidalgo2017simula}.
Further studies~\cite{parisi2005microscopic,parisi2007morphological} used the social force model to study the effect of desired velocity and the exit door on the duration of clogs.
Clogs caused by higher desired velocity in force-based models result in lower flow through bottlenecks, a phenomenon also known as ``faster-is-slower''~\cite{helbing2005self, parisi2005microscopic, parisi2007morphological, pastor2015experiment, patterson2017clogging}.

In this paper, we focus on prolonged clogs occurring in the generalized collision-free velocity model (GCVM)~\cite{xu2019generalized}, a first-order microscopic model for pedestrian dynamics.
It is based on the collision-free speed model~\cite{tordeux2016collision}, and strictly follows the principle of volume exclusion to guarantee that there is no overlap among agents.
Therefore, clogs that result in long-term interruption of flow occur frequently in simulations of bottlenecks, particularly in narrow exits.
We aim to quantify these prolonged clogs by exploring decisive factors behind their occurrence in the bottleneck scenario, to purposefully improve the GCVM for reproducing pedestrians' movement more realistically.
The effect of three types of factors is examined in this study. 
The first category includes two parameters of the spatial boundaries, i.e., the width and the position of the bottleneck exit.  
The second category consists of algorithmic factors related to the implementation of the GCVM, including the time step size and the update scheme (e.g., sequential or parallel update) for the agents in the simulations.
Third, several model parameters such as the strength of impact among agents in the GCVM, and the shapes of the agents are analyzed. 
The results are used to ascertain the relationship between these factors and the occurrence of prolonged clogs.

This paper is organized as follows. 
Section~\ref{sec:setup} introduces the bottleneck scenario for the simulations.
In section~\ref{sec:definition}, we briefly define the GCVM and introduce the method used for identifying prolonged clogs in numerical simulations.
Section~\ref{sec:experiment} compares simulation results obtained with various factors and shows the corresponding analysis.
Finally, we conclude with a discussion in section~\ref{sec:conclusion}.

\section{The bottleneck scenario for simulations}
\label{sec:setup}
\begin{figure}[H]
    \centering
    \includegraphics[width=0.8\linewidth]{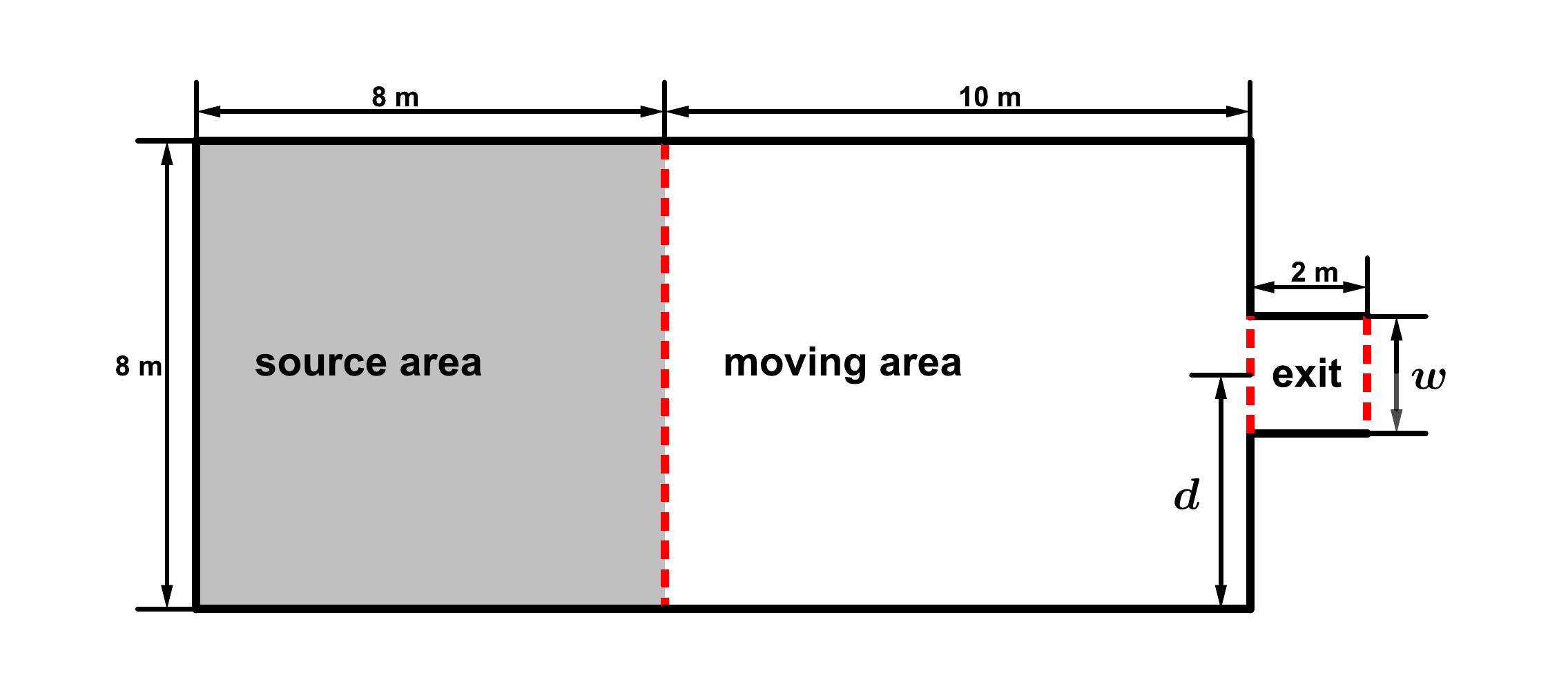}
    \caption{The bottleneck scenario for simulations.}
    \label{fig:geo}
\end{figure}

The bottleneck scenario for simulations in this study is shown in figure~\ref{fig:geo}.
It is composed of three parts separated by red dashed segments.
The source area, a $\SI{8}{\metre}\times\SI{8}{\metre}$ square in gray, the moving area, a rectangular room with an area of $\SI{10}{\metre}\times\SI{8}{\metre}$, and the exit, a corridor measuring $\SI{2}{\metre}\times w$.
In section~\ref{sec:experiment} different values of $w$ and $d$ (the position of the exit with respect to the lower horizontal wall) are used to determine the effect of the structure of the bottleneck has on the occurrence of the prolonged clogs.

In order to determine the decisive factors behind the appearance of prolonged clogs, simulations are implemented in the bottleneck scenario with different initial and boundary conditions.
In each simulation, 400 agents are generated with a constant rate at random positions in the source area and these move through the moving area to leave the scenario by the exit.
During this process, clogs may appear, leading to an interruption of the bottleneck flow.
A clog interrupting the flow longer than the time threshold $T_w$ is identified as a prolonged clog.
Since prolonged clogs can last a long time and so as to ensure that the blockage does not stop the dynamics of the system, resulting in an impractically long simulation time, we manually solve them by moving one of the agents involved in the clog to free space in the source area.
The details of this manual clog-solving procedure will be elaborated in the next section.
The number of prolonged clogs in each simulation is recorded.
Then the results of different simulations are compared to explore the relationship between these factors and the occurrence of prolonged clogs.

The model for pedestrian dynamics and the approach to identify clogs are presented in the following section.

\section{Introducing the model and identifying prolonged clogs}
\label{sec:definition}
We begin this section with a brief introduction to the GCVM, which is the model used in this study.
It is defined as
\begin{equation}
    \dot{X}_i(X_i,X_j,\dots)=\vec{e}_i(X_i,X_j,\dots)\cdot V_i(X_i,X_j,\dots),
\end{equation}
where $X_i$ is the position of agent $i$, $V_i$ is a scalar denoting its speed, and $\vec{e}_i$ is a unit vector representing its direction of movement.

\begin{figure}[H]
    \centering
    \subfigure[]{\includegraphics[width=0.45\linewidth]{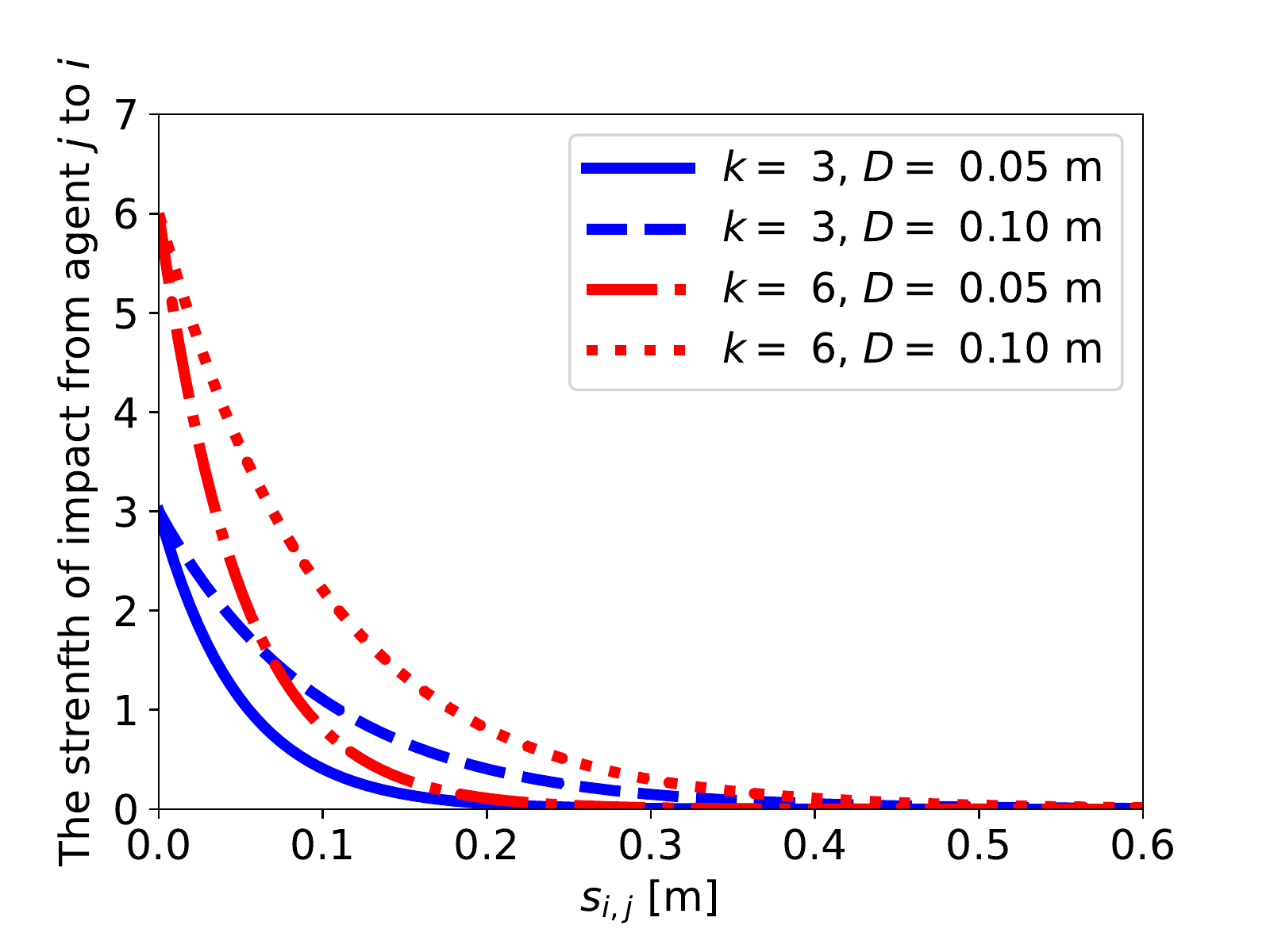}\label{fig:2a}}
    \subfigure[]{\includegraphics[width=0.45\linewidth]{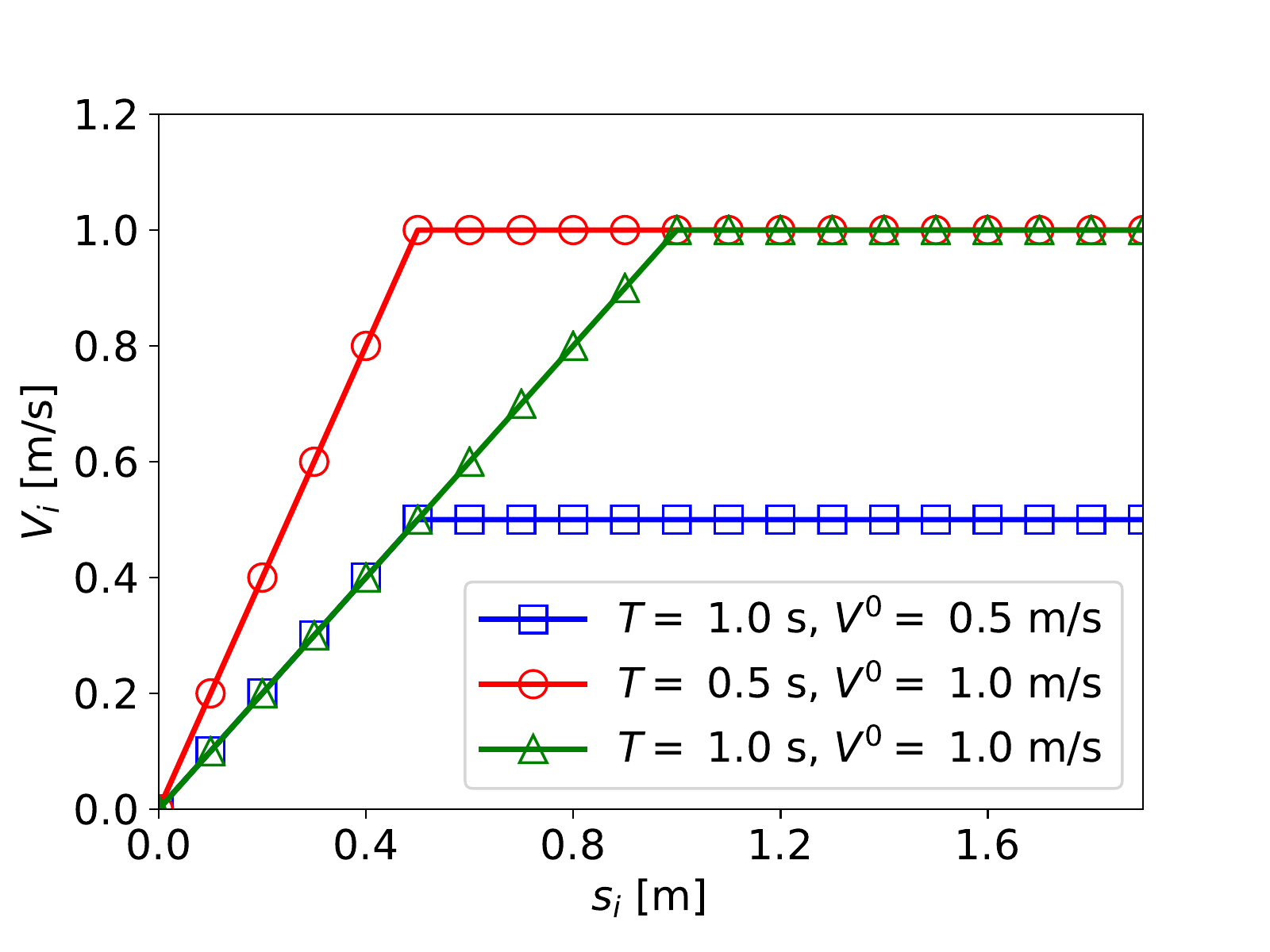}\label{fig:2b}}
    \caption{
    (a) The effect of $k$ and $D$ on the strength of impact.
    (b) The speed functions with different $V^0_i$ and $T$.
    \label{fig:ParaEffect}
    }
\end{figure}

The direction of movement $\vec{e}_i$ is calculated first by using the equation
\begin{equation}
\label{equ:2}
    \vec{e}_i=u \cdot\bigg(\vec{e}_i^{~0}+\sum_{j\in J_i} k\cdot \exp\Big(\frac{-s_{i,j}}{D}\Big)\cdot \vec{n}_{i,j}+\vec{w}_i\bigg).
\end{equation}
Here, $u$ is a normalization constant such that $\Vert\vec{e}_i\Vert=1$.
$\vec{e}_i^{~0}$ is a unit vector representing the desired direction of the agent.
This is calculated according to reference lines indicated by the red dashed segments in figure~\ref{fig:geo}.
The vector $\vec{e}_i^{~0}$ points to the middle of the reference line when agent $i$ is not in the range of the reference line; otherwise, it points to the nearest point on the reference line. 
More details of the calculation method are given in \cite{chraibi2012validated}.
$J_i$ is the set of agents that contains all neighbors affecting the moving direction of agent $i$.
The magnitude of the impact from these neighbors is a function of $s_{i,j}$, which is the distance between the edges of agent $i$ and $j$ along the line connecting their centers.
Parameters $k>0$ and $D>0$ are used to calibrate the strength and range of the impact, respectively.
The effect of $k$ and $D$ on the strength of impact is shown in figure~\ref{fig:2a} and a similar analysis for the effect of $k$ and $D$ can be found in \cite{hein2019agent}.
The direction of the impact from agent $j$ to $i$ is denoted by the unit vector $\vec{n}_{i,j}$, which depends on the relative positions of both agents.
$\vec{w}_i$ is the effect from walls and obstacles in the room, which is calculated analogously to the effect from neighbors.

Then the speed on the new moving direction is obtained using the equation 
\begin{equation}
\label{equ:3}
    V_i=\min\Big\{V^0_i,\max\big\{0,\frac{s_i}{T}\big\}\Big\}.
\end{equation}
The speed is a function of $s_i$, which is the maximum space of agent $i$ in the new direction of movement $\vec{e}_i$ without overlapping with other agents.
In equation~\eqref{equ:3}, $V^0_i$ is the free speed of agent $i$, the speed that is achieved by moving without interference from other agents. 
The parameter $T>0$ is the slope of the speed-headway relationship. 
The speed functions with different $V^0_i$ and $T$ are shown in figure~\ref{fig:2b}. 
The value of $T$ could be used to model the level of motivation in simulations.
A decrease of $T$ at constant $s_i$ leads to a smaller distance between agent $i$ and the nearest agent in front, which corresponds to behavior with a higher level of motivation.
A more detailed introduction to the GCVM can be found in~\cite{xu2019generalized}.

\begin{figure}[H]
    \centering
    \subfigure[]{\includegraphics[width=0.4\linewidth]{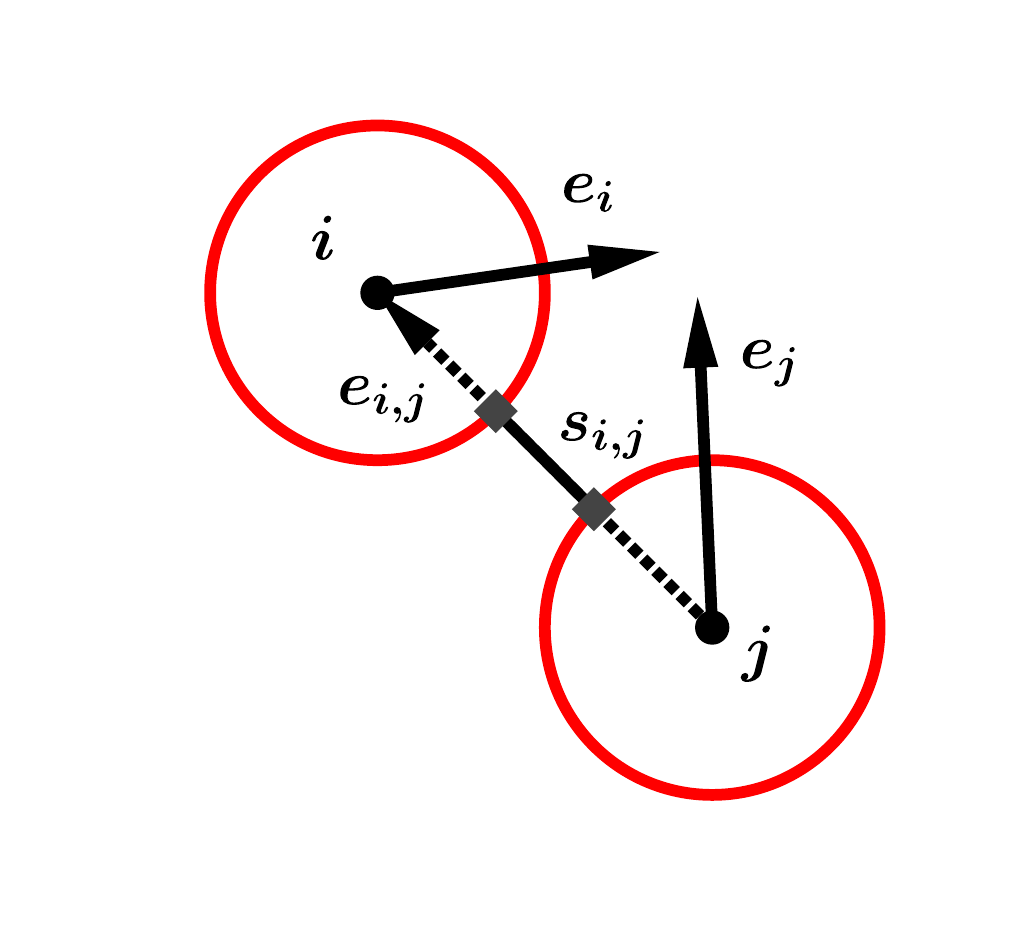}\label{fig:3a}}
    \subfigure[]{\includegraphics[width=0.25\linewidth]{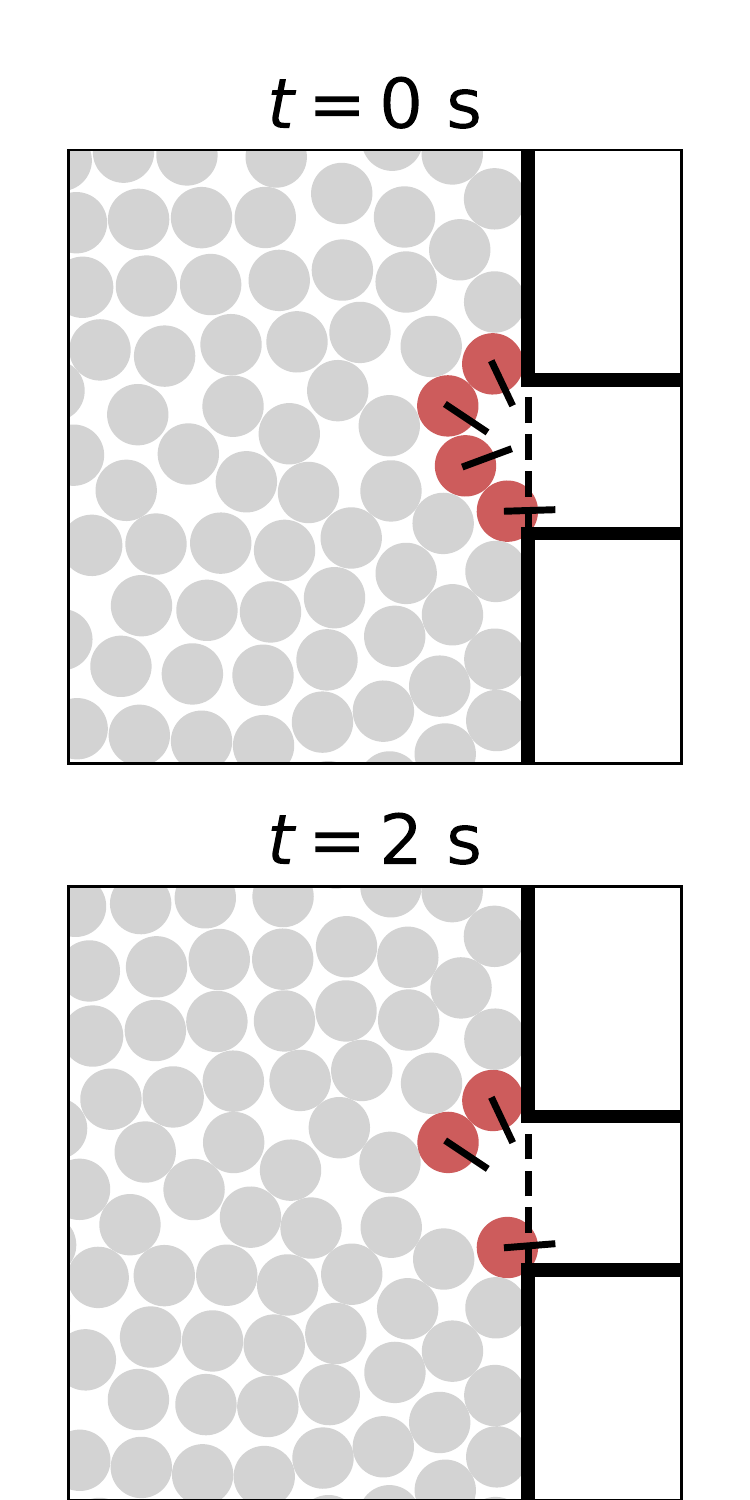}\label{fig:3b}}
    \subfigure[]{\includegraphics[width=0.25\linewidth]{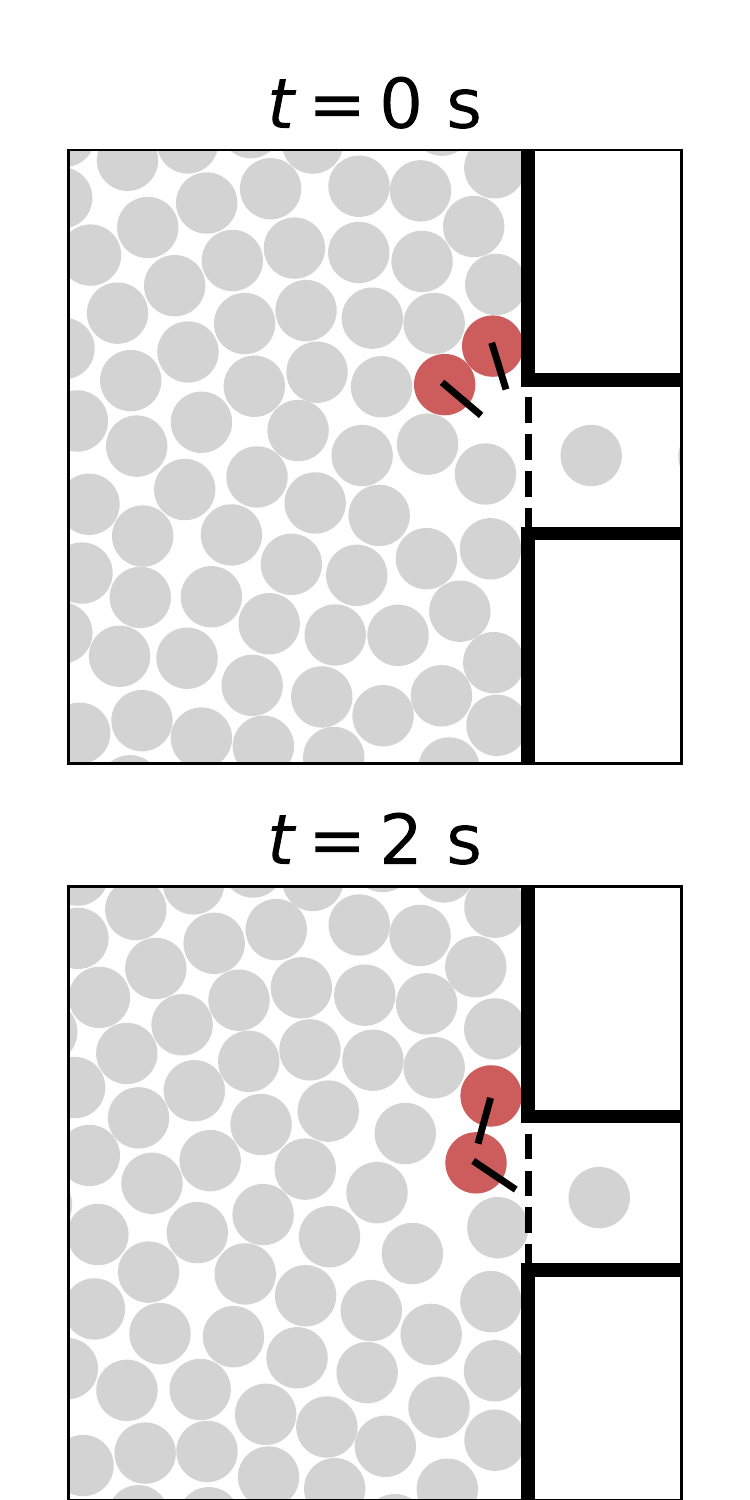}\label{fig:3c}}
    \caption{ 
    (a): When two agents are about to cause clogging,
    $\vec{e}_i$ and $\vec{e}_j$ are directions of movement of agent $i$ and $j$,
    $\vec{e}_{i,j}$ is the unit vector points from the center of agent $j$ to $i$,
    $s_{i,j}$ is the distance between the edges of agent $i$ and $j$ along the line connecting their centers.
    (b): A prolonged clog is manually solved after interrupting the flow for \SI{2}{\second}.
    (c): A clog is solved automatically by agents adjusting their direction of movement.
    }
    \label{fig:clogging}
\end{figure}

Since there is no overlapping among agents in the GCVM, the space occupied by one agent is not available to other agents.
Therefore, clogging occurs when the direction of movement, $\vec{e}_i$, of two agents point toward each other and the distance $s_{i,j}$ between them is too small for them to move.
A representative case is shown in figure~\ref{fig:3a}. 
It could be formalized by
\begin{equation}
\label{equ:4}
\begin{cases}
    s_{i,j}&\le \epsilon,\\
    V_i+V_j&\le \lambda, \\
    \vec{e}_{i,j}\cdot\vec{e}_i&<0,\\
    \vec{e}_{i,j}\cdot\vec{e}_j&>0,
\end{cases}
\end{equation}
where $\vec{e}_{i,j}$ is the unit vector points from the center of agent $j$ to $i$,
$\epsilon$ is a threshold used to determine whether the distance between these two agents is small enough to form a clog, 
and $\lambda$ is the threshold of speed to ascertain whether these two agents are almost stationary.
The last two conditions in equation~\eqref{equ:4} denote that these two agents are moving toward each other.
In the present study, $\epsilon$ is equal to the radius of agents, and $\lambda$ is set as $(V_i^0+V_j^0)/100$.
A clog formed by more than two agents contains at least two agents satisfying equation~\eqref{equ:4}.

There could be many pairs of agents that satisfy the definition of clogging in equation~\eqref{equ:4} at any time and any place in the simulation.
We treat clogs that interrupt the flow longer than time period $T_w$ as prolonged clogs.
These prolonged clogs occur almost around exits, as the degree of freedom in the direction of movement is limited by the wall.
An example of prolonged clogs is shown in figure~\ref{fig:3b}, a clog consisting of four red agents is formed in front of the bottleneck and interrupts the flow.
After $T_w=\SI{2}{\second}$, the clog is solved manually by moving one of the agents in the clog.
As for clogs that do not interrupt the flow or last less than $T_w$ seconds, we do not destroy them artificially since these can be solved automatically by agents adjusting their direction of movement.
A clog formed by two red agents, which can be automatically solved in $T_w=\SI{2}{\second}$, is shown in figure~\ref{fig:3c}.

\begin{figure}[H]
    \centering
    \includegraphics[width=0.85\linewidth]{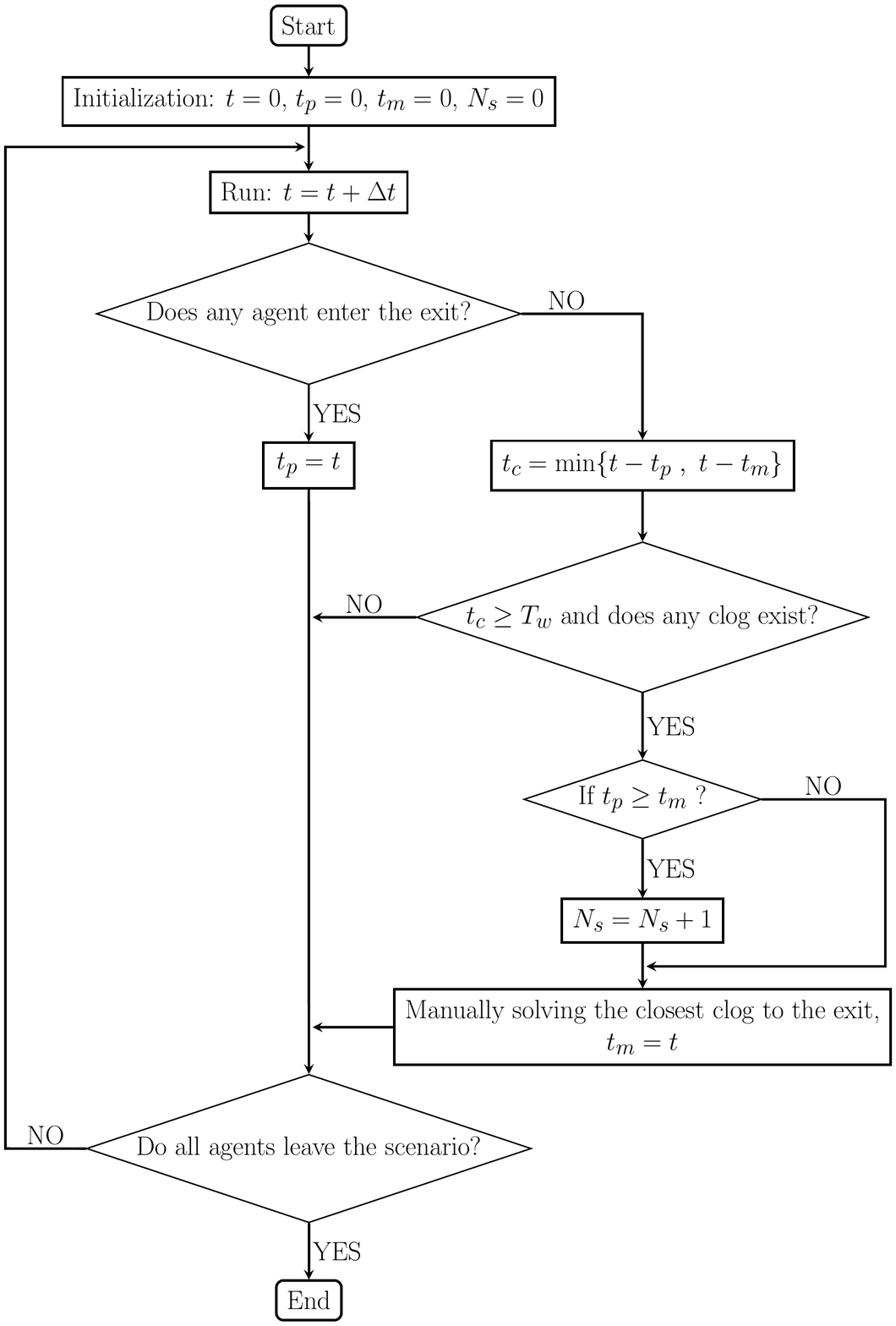}
    \caption{The process of solving and counting prolonged clogs.
    $t$ is the current time,
    $t_p$ is the time of the last agent entering the exit,
    $t_m$ is the time of the last manual clog-solving process,
    $N_s$ is the number of prolonged clogs,
    $\Delta t$ is the time step size in the simulation,
    $t_c$ is the smaller of $t-t_p$ and $t-t_m$,
    $T_w$ is the time threshold.}
    \label{fig:removepro}
\end{figure}

The flowchart in figure~\ref{fig:removepro} illustrates how to count the prolonged clogs in simulations, where $t$ is the current time, $t_p$ is the time at which the last agent enters the exit, $t_m$ is the time of the last manual clog-solving process, $N_s$ is the number of prolonged clogs, $\Delta t$ is the time step size in the simulation, and $t_c$ is the smaller of $t-t_p$ and $t-t_m$.

For each time step of a simulation, a non-zero flow through the measurement line between moving area and exit is an indicator that no prolonged clogs occur.
Otherwise, we will check whether $t_c$ is greater than the threshold $T_w$, and whether there are agents satisfying the definition of clogging in equation~\eqref{equ:4}.
A prolonged clog is identified if these two conditions are met.
It is treated as a new clog if $t_p$, the time when the last agent crossed the bottleneck, is not less than $t_m$, the time of the last manual removal of an agent.
Regardless of whether the clog is new or already existing, one of the two agents forming the closest clog to the exit is moved manually to free space in the source area.
It should be noted that breaking up a prolonged clog may require more than one manual clog-solving process, which results in $t_p < t_m$.
The number of prolonged clogs is counted from the beginning of the simulation to the last agent leaving the simulation scenario.

\section{Simulation results}
\label{sec:experiment}
In each simulation, one or two factors were selected for variation. 
The other factors were set to default values as shown in table ~\ref{tab:defaultPara}.

\begin{table}[H]
    \centering
    \caption{Default values of factors in simulations. $w$ is the width of the exit, $d$ is the distance between the center of the exit and the lower horizontal wall of the moving area, $k$ and $D$ are parameters used to calibrate the strength and range, respectively, of the impact from neighbors in the movement direction, $V^0_i$ is the free speed, and $T$ is the slope of the speed-headway relationship.}
    \label{tab:defaultPara}
    \setlength{\tabcolsep}{3pt}
    \begin{tabular}{cc}
    \hline
    Factors & Default values\\
    \hline
    Agent generation rate  & 8 Agents/$\SI{}{\second}$ \\
    Agent shape & circle ($r=\SI{0.2}{\meter}$) \\
    Update method & parallel update \\
    Time step size $\Delta t$ & $\SI{0.05}{\second}$ \\
    $w$ (figure~\ref{fig:geo}) & $\SI{0.8}{\meter}$ \\
    $d$ (figure~\ref{fig:geo}) & $\SI{4}{\meter}$ \\
    $k$ (equation~\eqref{equ:2}) & 3 \\
    $D$ (equation~\eqref{equ:2}) & $\SI{0.1}{\meter}$ \\
    $V_i^0$ (equation~\eqref{equ:3}) & $\SI{1.34}{\meter\per\second}$ \\
    $T$ (equation~\eqref{equ:3}) & $\SI{0.3}{\second}$ \\
    \hline
    \end{tabular}
\end{table}

\begin{figure}[H]
    \centering
    \includegraphics[width=0.6\linewidth]{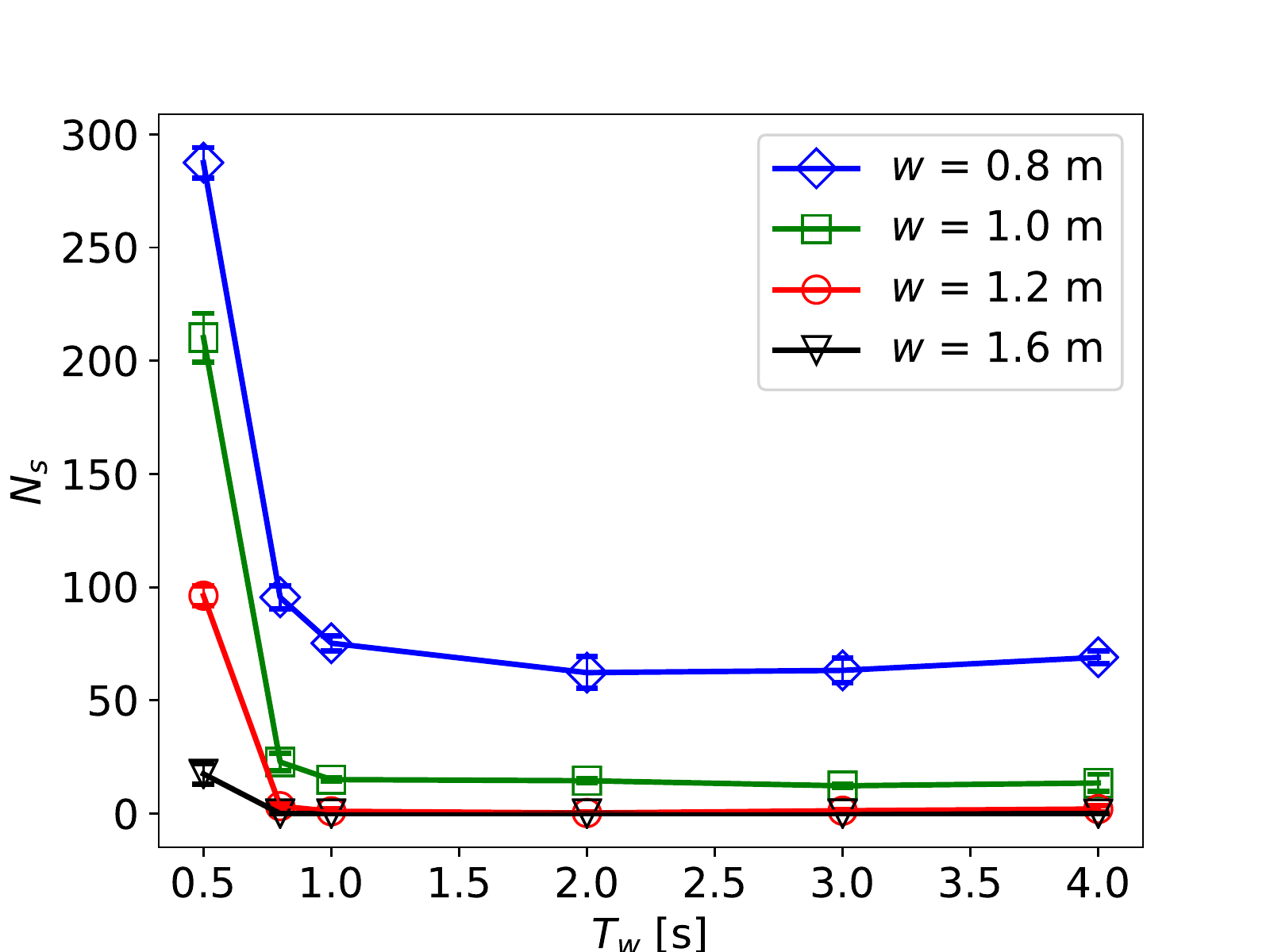}
    \caption{The correlation between $N_s$ (the number of prolonged clogs) and $T_w$ (the time span between the formation and artificial termination of a prolonged clog) for different values of $w$ (the width of the exit).
    The error bars show the standard deviations.}
    \label{fig:findTw}
\end{figure}

To improve the efficiency of simulations, a series of simulations were implemented firstly to select the suitable $T_w$, the time span between the formation, and artificial termination of a prolonged clog for subsequent simulations.
We ran simulations in four bottleneck scenarios, where the value of $w$ was 0.8, 1.0, 1.2,  and 1.6~$\SI{}{\meter}$, respectively.
For each scenario, simulations with $T_w$ from $\SI{0}{\second}$ to $\SI{4}{\second}$ were implemented.
We ran each simulation four times with different distributions of agents in the source area.
The relationship between $T_w$ and the mean values of $N_s$, the number of prolonged clogs, from the four runs are shown in figure~\ref{fig:findTw}, where the error bars indicate the standard deviations.
The results in scenarios with a different value of $w$ are represented by different marks and colors.
In all four scenarios, $N_s$ dis not change significantly when $T_w$ was longer than $\SI{2}{\second}$. 
Therefore, this value for $T_w$ was selected for the following simulations.

In the following, we ran each simulation four times.
The mean value of $N_s$ from the four runs reflects the effect of the factors observed on the occurrence of prolonged clogs. 
Moreover, the time lapse between two consecutive agents entering the exit, and the trajectories of agents were analyzed for the selected factors. 

\subsection{Parameters of spatial boundaries}
\label{sec:spatial}
The effects of the width and the position of the exit are explored in this subsection.
Three exit positions ($d=$ 4.0, 2.0, or $w/2$ $\SI{}{\meter}$) and six widths ($w=$ 0.8, 1.0, 1.2, 1.6, 2.0, or 2.5 $\SI{}{\meter}$) were selected for the simulations.
The exit was located in the middle of two lateral walls of the moving area when $d=\SI{4}{\meter}$ and adjacent to the lower horizontal wall when $d=w/2$. 

\begin{figure}[H]
    \centering
    \includegraphics[width=0.6\linewidth]{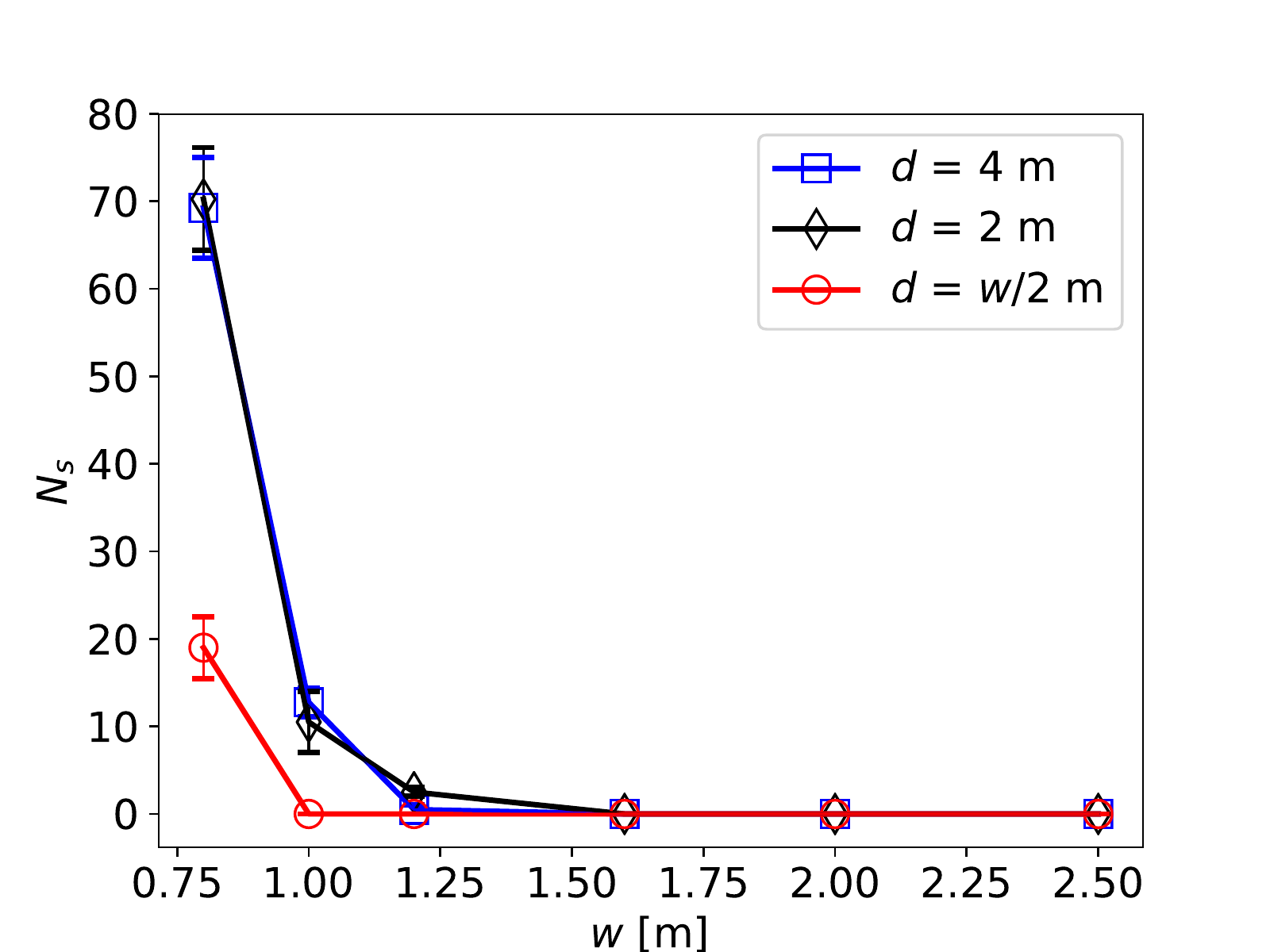}
    \caption{The correlation between $N_s$ (the number of prolonged clogs) and $w$ (the width of the exit) for different values of $d$ (the distance between the center of the exit and the lower horizontal wall of the moving area). 
    The error bars show the standard deviations.}
    \label{fig:6}
\end{figure}

Figure~\ref{fig:6} shows the correlation between $N_s$ and $w$ for different values of $d$.
The position of the exit does not alter the fact that $N_s$ decreases to zero as $w$ increases.
Moreover, there is no prolonged clog when the exit is wider than $\SI{1.6}{\meter}$ for all three positions.
Besides the effect of $w$, when $d=w/2$ (the exit is adjacent to the lower horizontal wall of the moving area), $N_s$ was significantly less than with the other two locations.
We assumed that this difference was caused by the reduced degree of freedom in the possible directions in which agents will move.

\begin{figure}[H]
    \centering
    \subfigure[]{\includegraphics[width=0.45\linewidth]{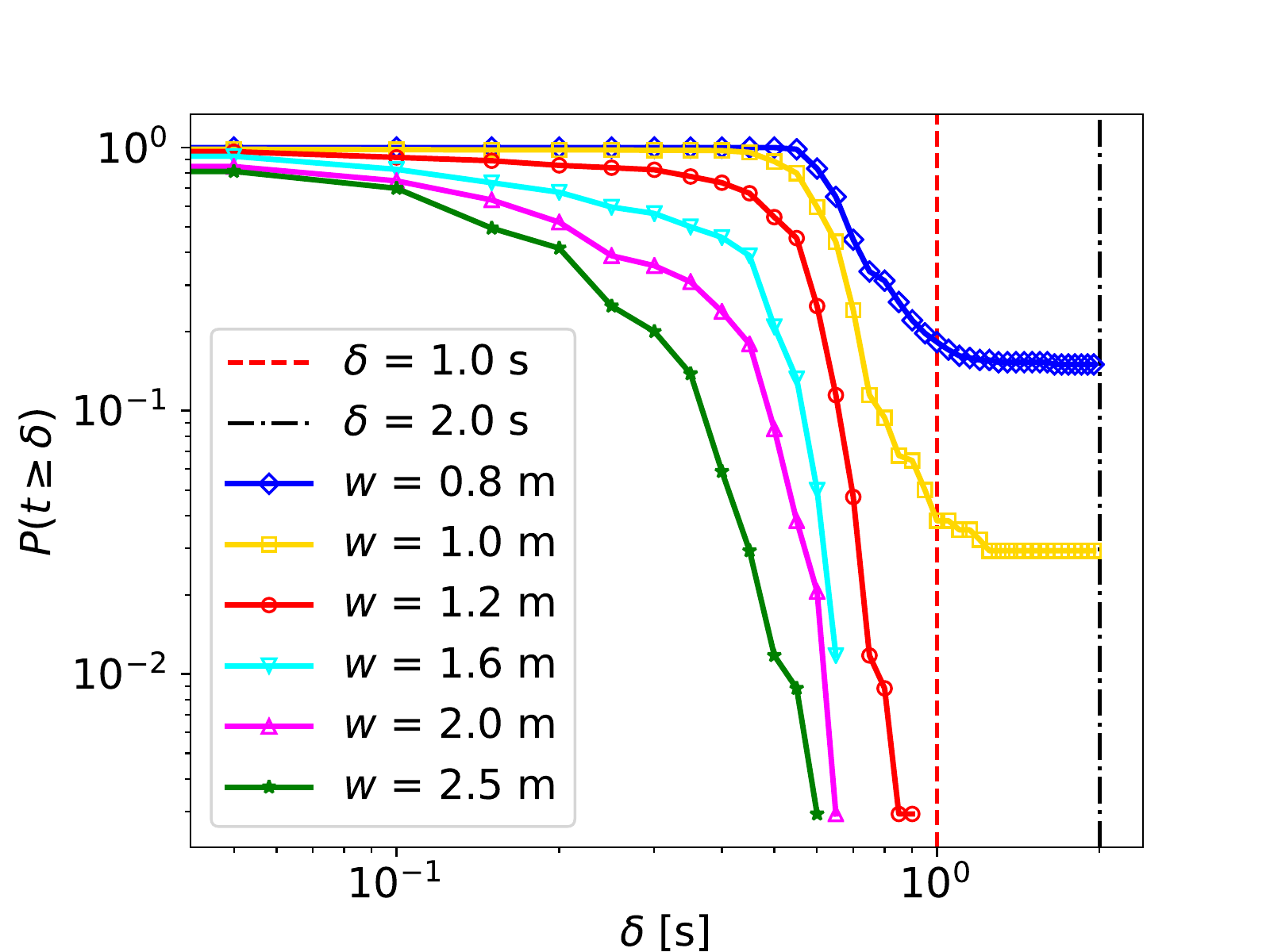}\label{fig:7a}}
    \subfigure[]{\includegraphics[width=0.45\linewidth]{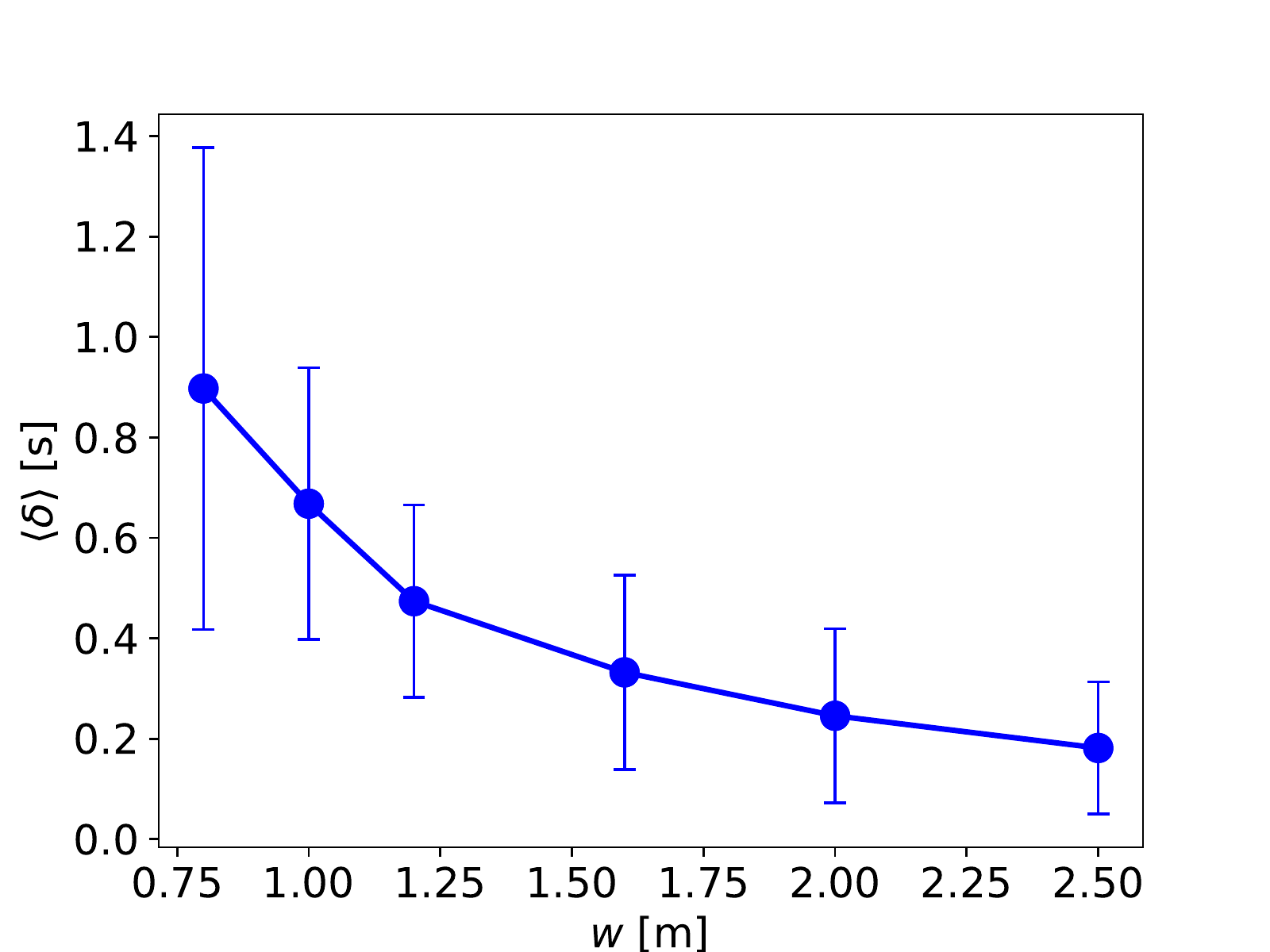}\label{fig:7b}}
    \caption{
    (a): The survival functions of $\delta$ for simulations with different values of $w$ when $d = \SI{4}{\meter}$.
    (b): The correlation between the $\langle\delta\rangle$ (the mean time lapse between two consecutive agents entering the exit) and $w$ when $d = \SI{4}{\meter}$.
    The error bars show the standard deviations.
    }
\end{figure}

In order to quantitatively analyze the influence of  the width of the exit ($w$) on the clogs, we examined the time lapses $\delta$ between two consecutive agents passing the exit, for different values of $w$.
The value of $\delta$ reflects the sustained time of clogs interrupting the flow. 
The probability distribution function $P(t>\delta)$, also known as the survival function, is sensitive to changes in the spatial boundaries, e.g. the width of the bottleneck \cite{zuriguel2014clogging,garcimartin2015flow,garcimartin2016flow,garcimartin2014experiment}.
We analyzed the results of simulations when $d = \SI{4}{\meter}$.
The survival functions of different values of $w$ are compared in figure~\ref{fig:7a}.
It can be observed that the probability of a higher value of $\delta$ decreases as $w$ increases.
Besides, the occurrence of prolonged clogs leads to plateaus in the survival functions of $w = 0.8$ and $w = \SI{1.0}{\meter}$.
Basically, in these two cases, the actual values of $\langle\delta\rangle$, the mean time lapse, are unknown as clogs lasting longer than $\SI{2}{\second}$ are manually solved. 
In fact, the actual value of $\langle\delta\rangle$ without manually removal of clogs may probably tend to infinite.
Nevertheless, in order to obtain an estimate for the lower bound of $\langle\delta\rangle$ and study its dependence on $w$, we treated all $\delta>\SI{2}{\second}$ as $\delta=\SI{2}{\second}$ in the calculation of the mean value of $\delta$. 
The correlation between the value of $\langle\delta\rangle$ and $w$ is shown in figure~\ref{fig:7b}.
The mean value and standard deviation of $\delta$ both decrease with increasing $w$.

\subsection{Algorithmic factors}
The effect of update methods and the time step sizes $\Delta t$ to solve the equation of motion are analyzed in this subsection.
Two update methods were adopted: the parallel update and the sequential update.
When we used the parallel update, the direction of movement, speed and location of all the agents were updated at the same time.
When the sequential update was used, the direction of movement, speed and location of agents were updated one by one according to the distance to the exit.
The agents near the exit had more effect on the dynamic of the system than the agents further away from the exit. 
Therefore, the agent with a greater effect, i.e., the agent closer to the exit, was updated first in the sequential update.
For each update method, simulations were performed with different values of $\Delta t$ from $\SI{0.01}{\second}$ to $\SI{0.125}{\second}$.

\begin{figure}[H]
    \centering
    \includegraphics[width=0.6\linewidth]{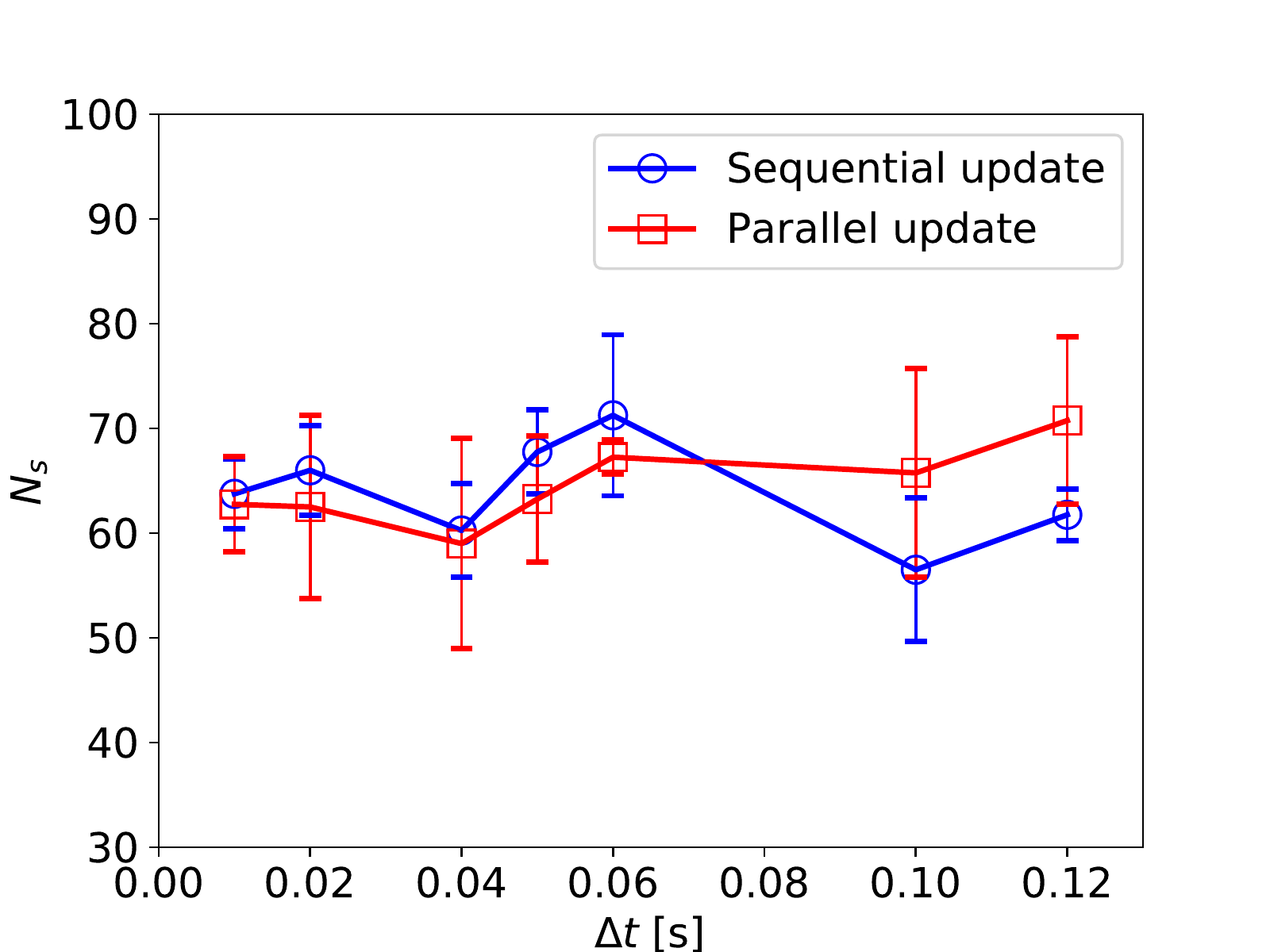}
    \caption{ The correlation between $N_s$ (the number of prolonged clogs) and $\Delta t$ (time step size) for different update methods.
    The error bars show the standard deviations.}
    \label{fig:technical}
\end{figure}

The correlation between $N_s$ and $\Delta t$ for two update methods is shown in figure~\ref{fig:technical}.
The effect of $\Delta t$ on $N_s$ is marginal for both update methods.

To explain the reason behind the results, an extreme case is considered here where two agents $i$ and $j$ are moving directly toward each other, which means $\vec{e}_i=-\vec{e}_j$.
We assume that their direction of movement is fixed, and $s_{i,j}<T \cdot V^0_i$.
According to equation~\eqref{equ:3}, their speeds $V_i$ and $V_j$ are both equal to $s_{i,j}/T$.
They will not overlap and, consequently, form a clog if their speeds satisfy 
\begin{equation}
    (V_i+V_j)\cdot \Delta t\le s_{i,j},
\end{equation}
which can be transformed to $2\cdot \Delta t \le T$.
This example illustrates that adopting a lower value of $\Delta t$ or substituting the sequential update cannot hinder the occurrence of clogging, since the scarcity of available space is not changed.

Therefore, the occurrence of prolonged clogs in the simulations with the GCVM is not an algorithmic issue.

\subsection{Parameters of the GCVM}
\label{subsec:Model_factor}
In this subsection, the effect of several parameters in the GCVM is examined, including the slope of the speed-headway relationship $T$ and the free speed $V^0$ in equation~\eqref{equ:3}. 
The strength and range of the effect of neighbors in the direction of movement, $k$ and $D$ in equation~\eqref{equ:2}, and the shapes of agents are also studied.

\begin{figure}[H]
    \centering
    \subfigure[]{\includegraphics[width=0.45\linewidth]{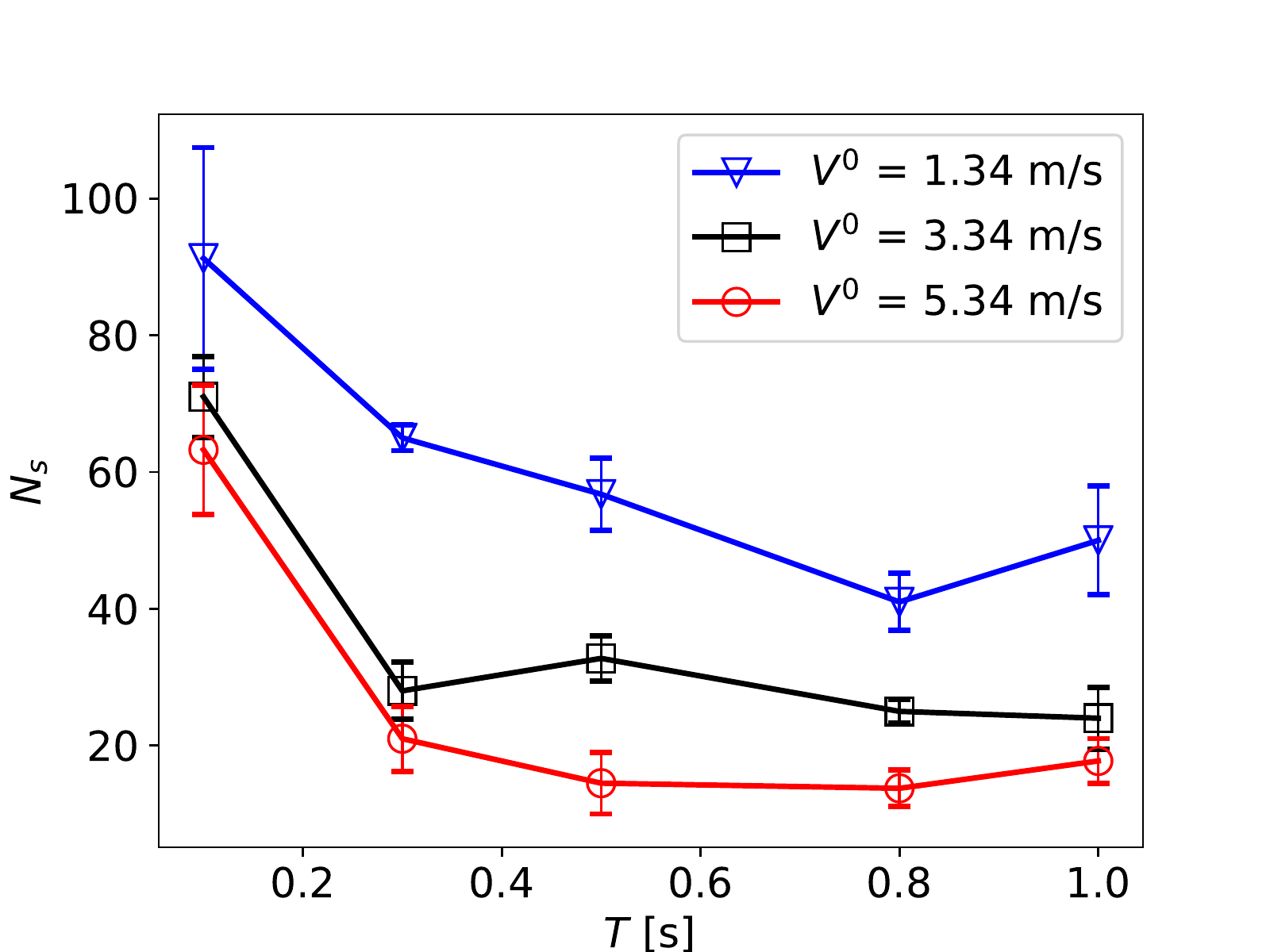}\label{fig:9a}}
    \subfigure[]{\includegraphics[width=0.45\linewidth]{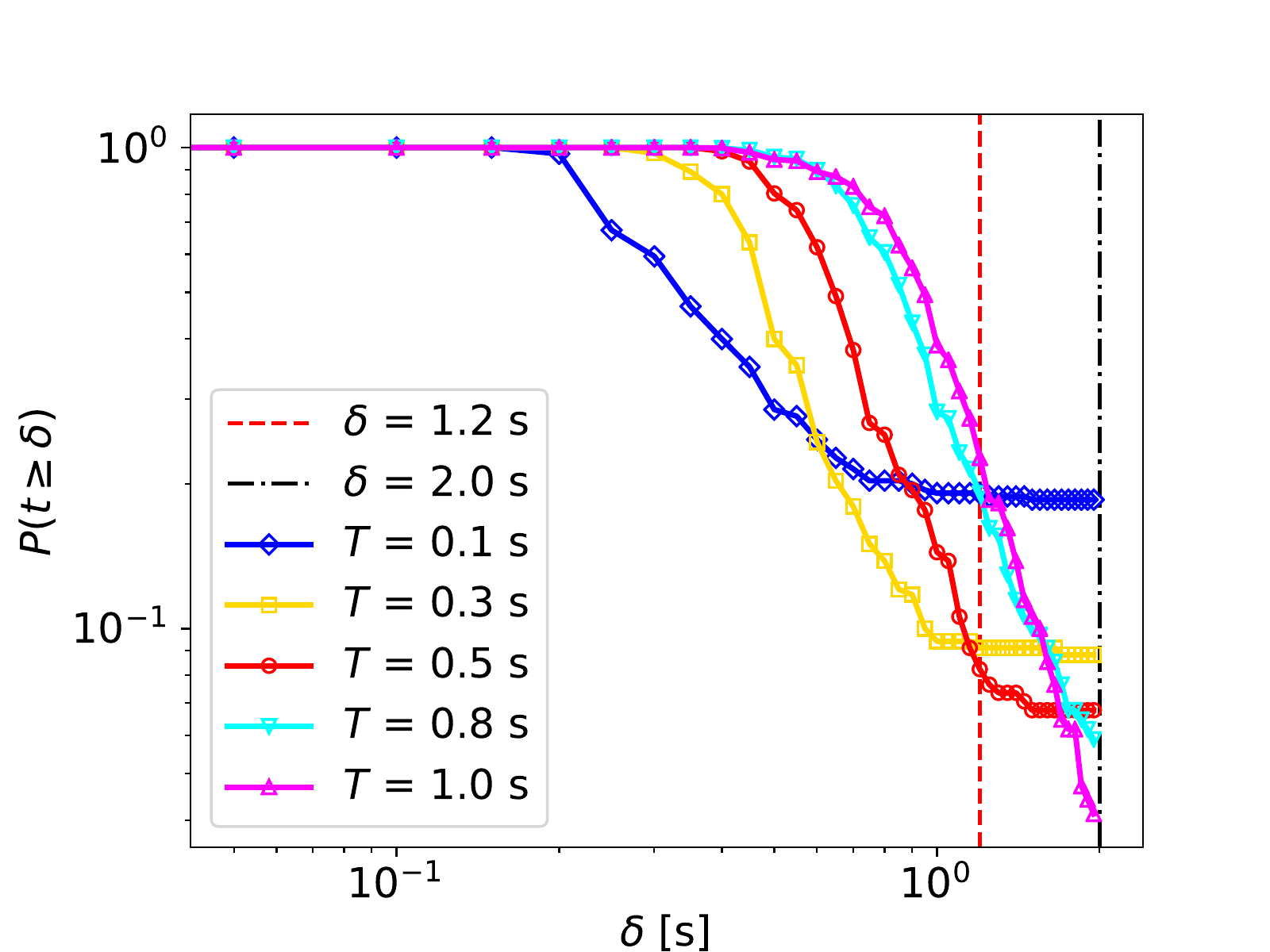}\label{fig:9b}}
    \subfigure[]{\includegraphics[width=0.45\linewidth]{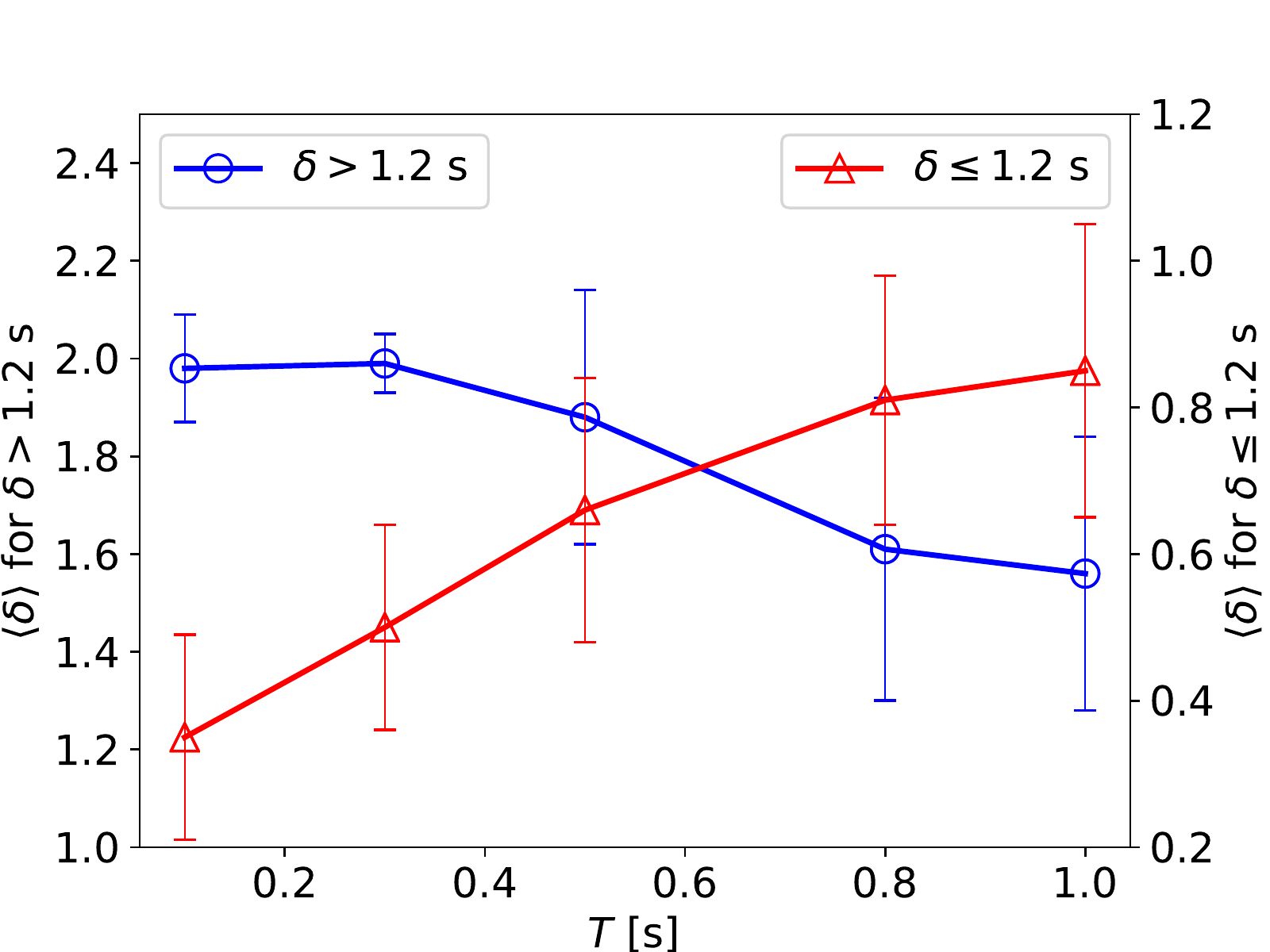}\label{fig:9c}}
    \subfigure[]{\includegraphics[width=0.45\linewidth]{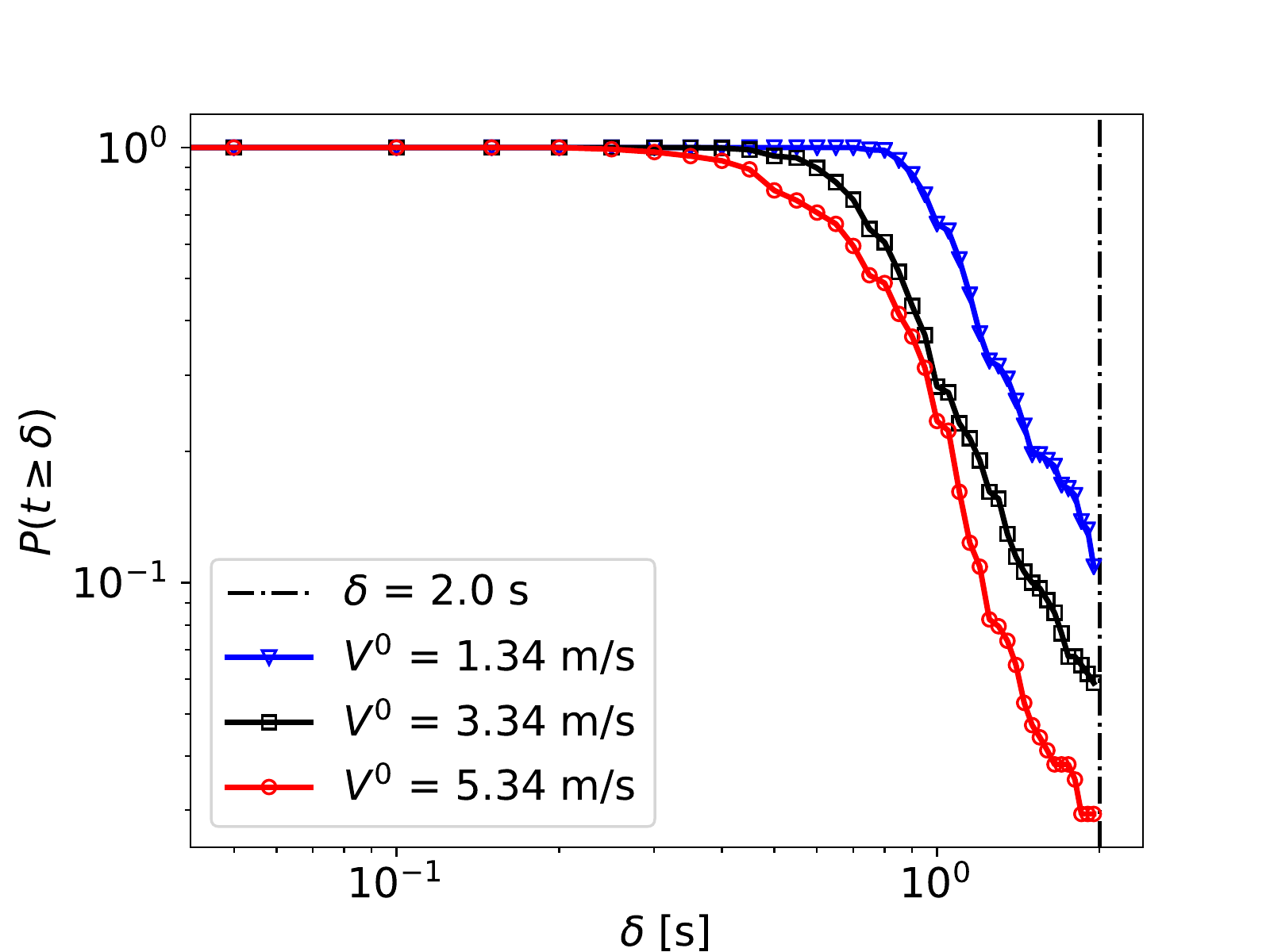}\label{fig:9d}}
    \caption{
    (a): The correlation between $N_s$ (the number of prolonged clogs) and $T$ (the slope of the speed-headway relation) for different values of $V^0$ (the free speed).
    The error bars show the standard deviations.
    (b): The survival functions of $\delta$ (the time lapse between two consecutive agents entering the exit) in the simulations with different values of $T$ when $V^0=\SI{3.34}{\meter\per\second}$.
    (c): The mean time lapse $\langle\delta\rangle$ versus $T$ when $V^0=\SI{3.34}{\meter\per\second}$, for $\delta \leq \SI{1.2}{\second}$ and $\delta>\SI{1.2}{\second}$, respectively.
    (d): The survival functions of $\delta$ (the time lapse between two consecutive agents entering the exit) in the simulations with different values of $V^0$ when $T=\SI{0.8}{\second}$.}
    \label{fig:TandV0}
\end{figure}

First, we looked at the effect of $T$ and $V^0$.
We ran simulations with different $V^0$ (1.34, 3.34, or 5.34 $\SI{}{\meter\per\second}$) and different $T$ (0.1, 0.3, 0.5, 0.8, or 1.0 $\SI{}{\second}$).
The correlation between $N_s$ and $T$ for different values of $V^0$ is shown in figure~\ref{fig:9a}.
For all three values of $V^0$, as $T$ increased, $N_s$ decreased initially, then remained relatively stable.
A decrease in $T$ led to a smaller slope of the speed-headway function, see equation~\eqref{equ:3} and figure~\ref{fig:2b}.
With decreasing $T$ agents move closer, which reduces the space available to resolve clogs.
This is in accordance with the finding that clogging is more likely to occur in scenarios with higher level of motivation \cite{garcimartin2016flow, hidalgo2017simula, pastor2015experiment, garcimartin2014experiment}.

The level of motivation has been shown to have an effect on the time lapse $\delta$ \cite{garcimartin2016flow, hidalgo2017simula, pastor2015experiment}.
Figure ~\ref{fig:9b} shows the survival functions of $\delta$ in the simulations with $V^0=\SI{3.34}{\meter\per\second}$, which is similar to the result of granular media experiment \cite{pastor2015experiment}.
These survival functions can be approximately separated into two successive regimes by $\delta=\SI{1.2}{\second}$. 
For $\delta \leq \SI{1.2}{\second}$, increasing $T$ leads to a higher value of $P(t>\delta)$, while for $\delta>\SI{1.2}{\second}$, increasing $T$ reduces $P(t>\delta)$.
The mean time lapse $\langle\delta\rangle$ for each regime is shown in figure~\ref{fig:9c}.
As we mentioned above, the actual values of $\langle\delta\rangle$ for the region of $\delta>\SI{1.2}{\second}$ are unknown as clogs lasting longer than $\SI{2}{\second}$ are manually solved.
Therefore, we treated all $\delta>\SI{2}{\second}$ as $\delta=\SI{2}{\second}$ in the calculation of the mean value of $\delta$.
The obtained values are the lower bound of the real ones.
Decreasing $T$ can be interpreted as increasing the level of motivation, which results in an increase in the free flow rate ($\delta \leq \SI{1.2}{\second}$) as well as an increase in the probability of clogging ($\delta>\SI{1.2}{\second}$).

However, a higher value of $V^0$, which can be interpreted as the expression of a higher motivation level, leads to lower values of $N_s$.
We analyzed the results of simulations with $T=\SI{0.8}{\second}$.
The survival functions of different values of $V^0$ are compared in figure~\ref{fig:9d}.
The probability of a higher value of $\delta$ decreases as $V^0$ increases.
According to equation~\eqref{equ:3}, the speed of agents in the GCVM depends on the overlapping-free spaces in their directions of movement. 
Although a higher $V^0$ increases the maximum possible speed of agents, it has little effect in congested areas due to limited space.
Therefore, the effect of $V^0$ in the GCVM on the motivation level of present simulations is marginal, as most of the investigated situations represent congested conditions.
Moreover, a higher $V^0$ allows agents to move faster in low density situations, which results in the reduction of $N_s$.
Note, that in force-based models~\cite{hidalgo2017simula} the driving force increases with increasing $V^0$, hence $V^0$ can have an effect in congested situations as well.

\begin{figure}[H]
    \centering
    \includegraphics[width=0.6\linewidth]{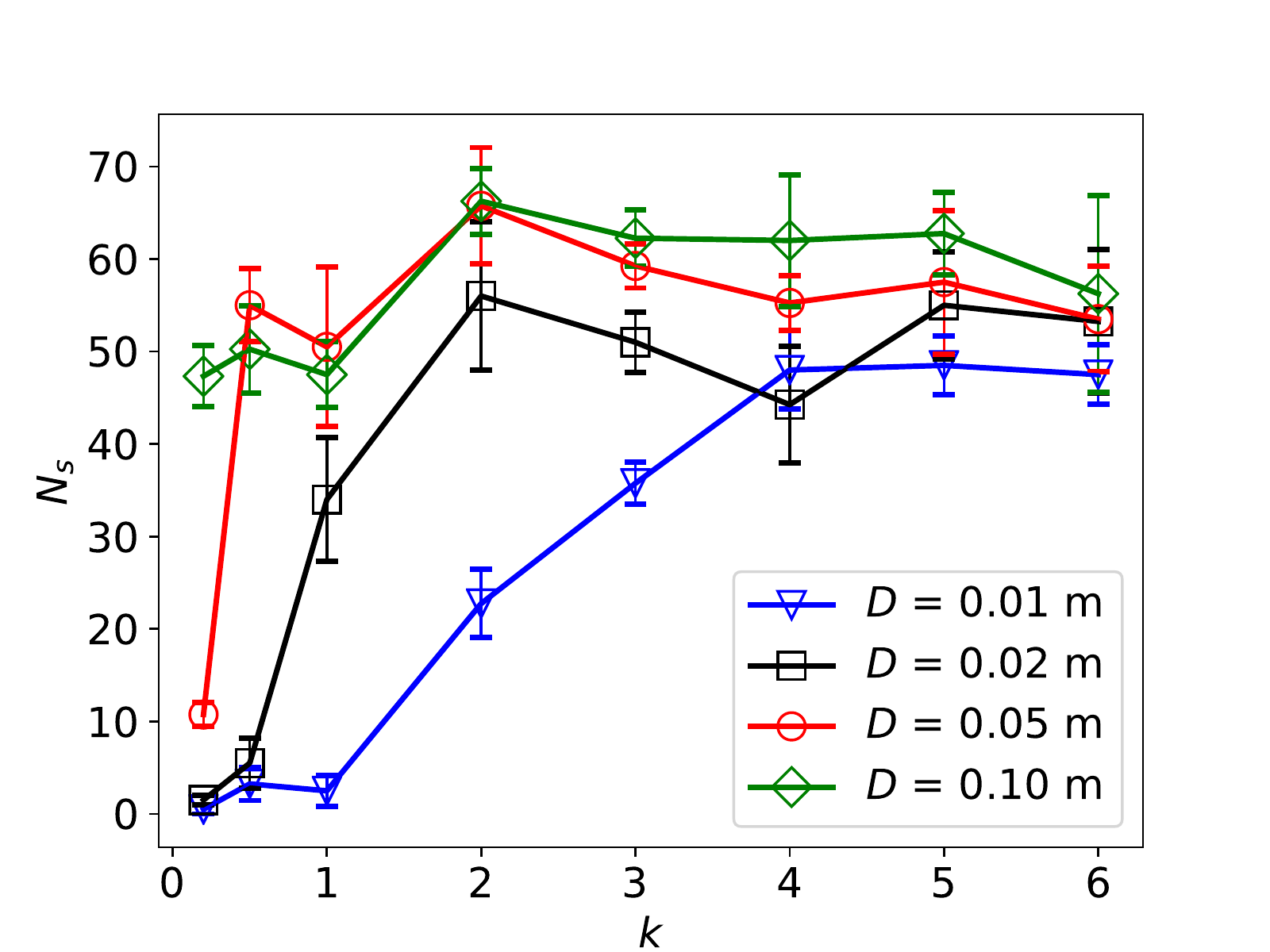}
    \caption{
    The correlation between $N_s$ (the number of prolonged clogs) and $k$ for different values of $D$.
    The error bars show the standard deviations.
    }
    \label{fig:10}
\end{figure}

Then we examined the effect of $k$ and $D$. 
Higher values of $k$ and larger $D$ led to agents being more stimulated to deviate from their desired directions.
We ran simulations with different values of $k$ (0.2, 0.5, 1, 2, 3, 4, 5, or 6) and different values of $D$ (0.01, 0.02, 0.05, or 0.1~$\SI{}{\meter}$).
The correlation between $N_s$ and $k$ for different values of $D$ is shown in figure~\ref{fig:10}.

\begin{figure}[H]
    \centering
    \subfigure[]{\includegraphics[width=0.45\linewidth]{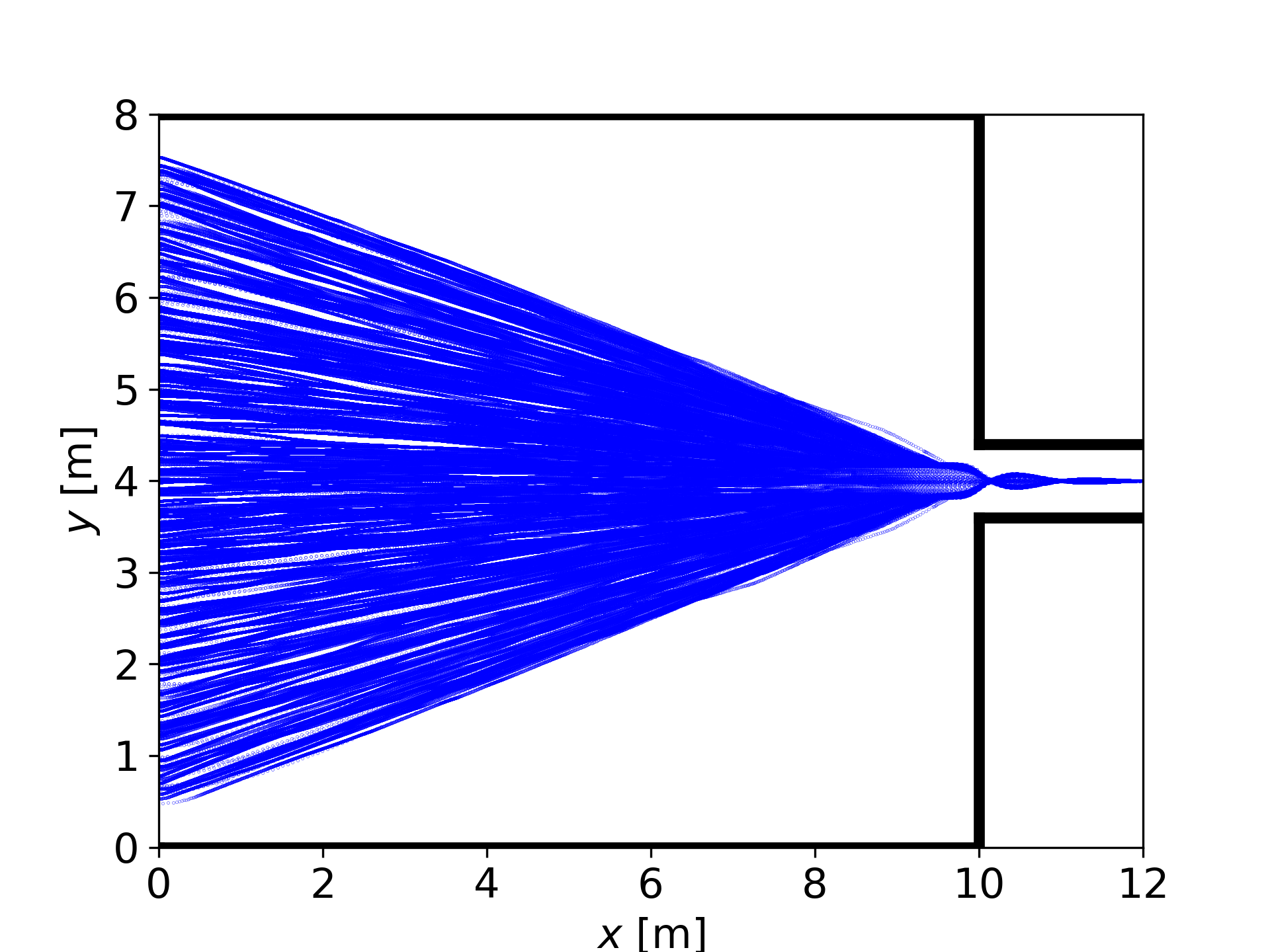}\label{fig:11a}}
    \subfigure[]{\includegraphics[width=0.45\linewidth]{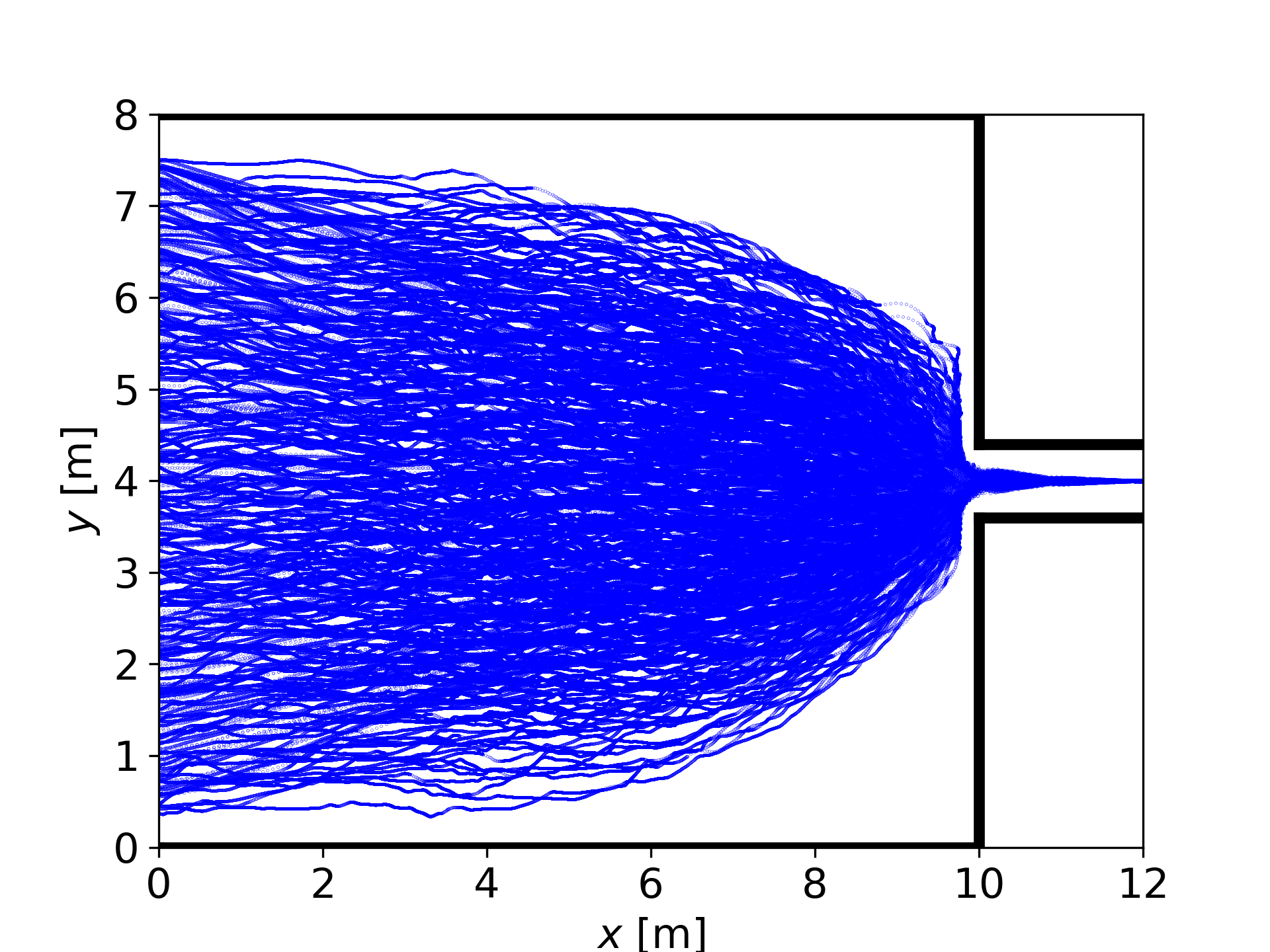}\label{fig:11b}}
    \caption{
    (a): Trajectories of agents when $k$ is 0.2 and $D$ is $\SI{0.01}{\meter}$.
    (b): Trajectories of agents when $k$ is 3.0 and $D$ is $\SI{0.01}{\meter}$.
     }
    \label{fig:kandD}
\end{figure}

It can be seen that $N_s$ increases with increasing $k$ and increasing $D$.
We assume the reason for this is that lower values of $k$ and $D$ decrease the neighbor's impact on agents, which leads to the queuing behavior.
We show in figure~\ref{fig:11a} the trajectories of agents in the simulation when $k$ is 0.2 and $D$ is $\SI{0.01}{\meter}$, which shows a strong queuing behavior.
When the impact among agents increased consistently, agents began to deviate from their desired direction until the queuing behavior was broken.
Figure~\ref{fig:11b} shows the trajectories of agents in the simulation when $k$ is 3 and $D$ is $\SI{0.01}{\meter}$.

The final factor analyzed was the shapes of agents.
In the previous sections, a pedestrian's shape was modeled as circles with a constant radius. 
To study the influence of the shape, we also performed simulations where pedestrians were modeled as velocity-based ellipses~\cite{xu2019generalized}.
The length of the semi-axis along the walking direction is a constant value $a$. 
The length of the other semi-axis along the shoulder equals $b$, which is defined as
\begin{equation}
    b=b_{\min}+\frac{b_{\max}-b_{\min}}{1+e^{\beta \cdot(V-\gamma)}}~,
\end{equation}
where $b_{\max}$ is the maximum value which is equal to half of a static pedestrian's width, $b_{\min}$ is equal to the half of a moving pedestrian's minimum width, $V$ is the speed of the agent, and parameters $\beta$ and $\gamma$ are used to adjust the shape of the function.

\begin{figure}[H]
    \centering
    \subfigure[]{\includegraphics[width=0.45\linewidth]{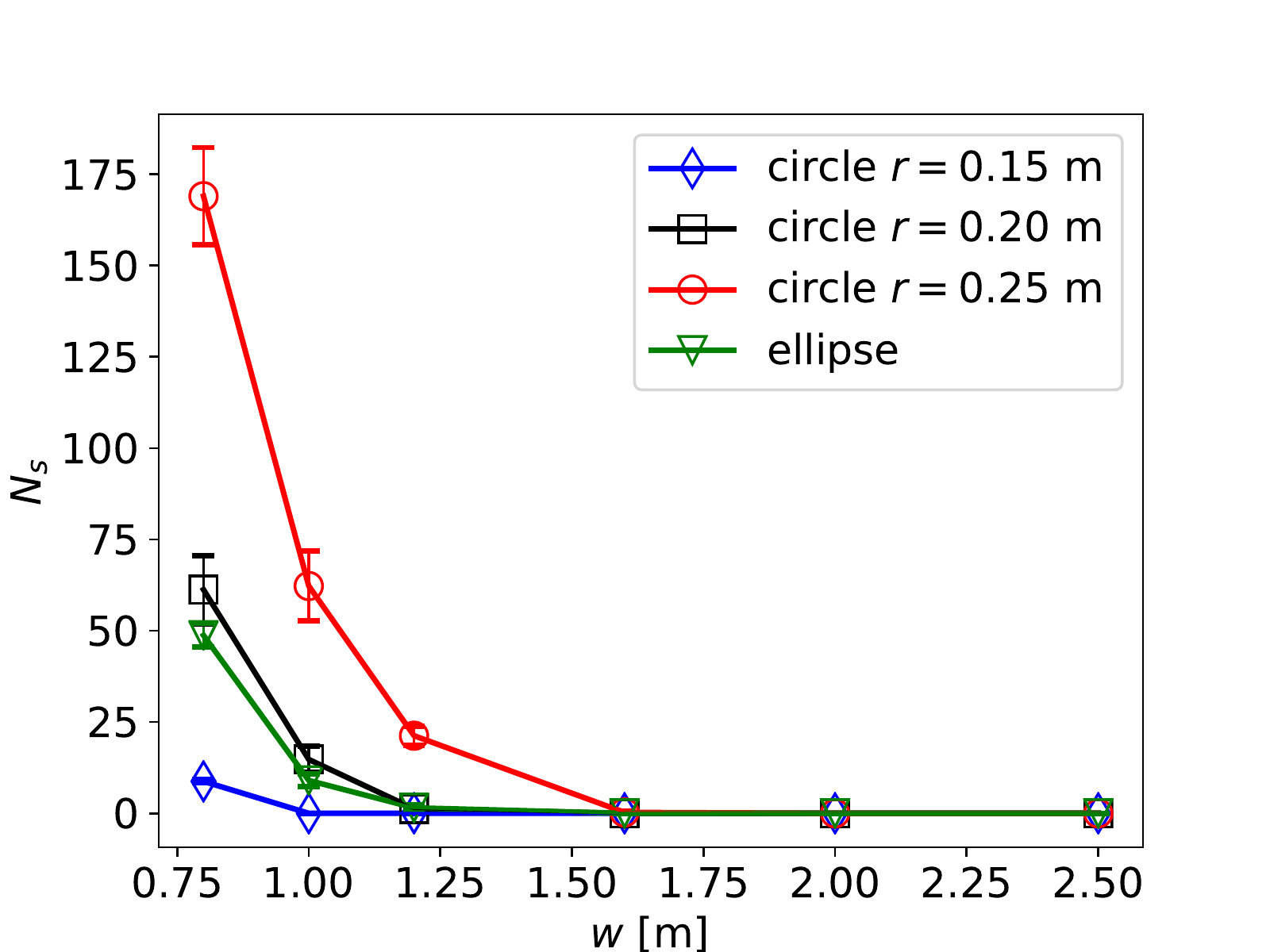}\label{fig:12a}}
    \subfigure[]{\includegraphics[width=0.45\linewidth]{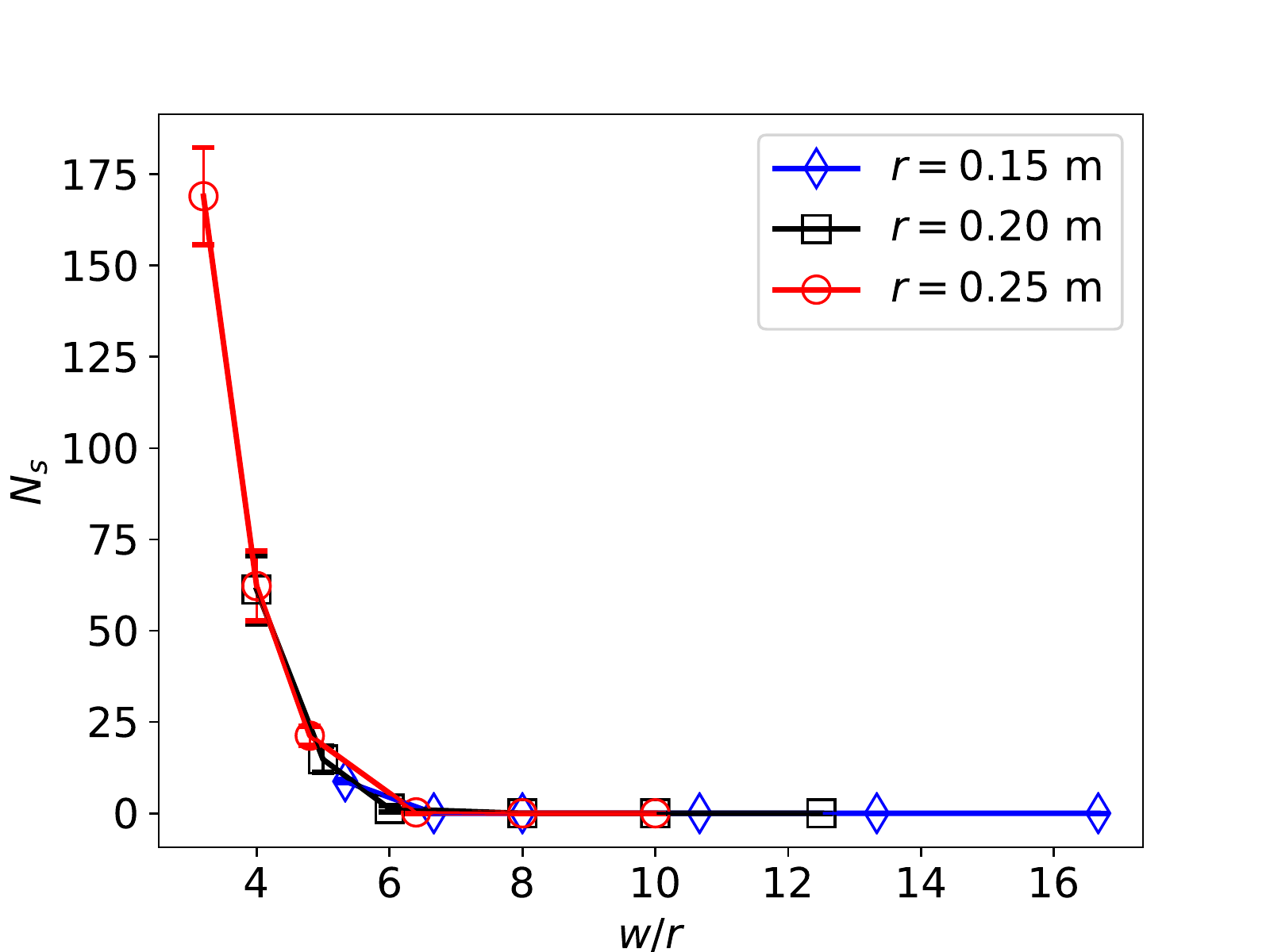}\label{fig:12b}}
    \caption{
    (a) The correlation between $N_s$ (the number of prolonged clogs) and $w$ (the width of the exit) for different shapes of agents.
    The error bars show the standard deviations.
    (b) The correlation between $N_s$ and $w/r$ for different $r$ (the radius of agents).
    The error bars show the standard deviations.}
    \label{fig:shapes}
\end{figure}

Simulations in this part are performed with three constant circles with different radius values $r$ (0.15, 0.20, or 0.25~$\SI{}{\meter}$) and a velocity-based ellipse ($a=\SI{0.20}{\meter}$, $b_{\min}=\SI{0.15}{\meter}$, $b_{\max}=\SI{0.25}{\meter}$, $\beta=50$, $\gamma=0.1$).
Figure~\ref{fig:12a} shows the correlation between $N_s$ and $w$ for different shapes.
The result of the ellipse is close to the result of the circle  with $r=\SI{0.20}{\meter}$, which is between the result of the smallest circle ($r=\SI{0.15}{\meter}$) and of the biggest circle ($r=\SI{0.25}{\meter}$).
The possible explanation of this result is that the shape of the dynamic ellipse varies with the speed of agents.
For example if the speed of an agent is $\SI{0.1}{\meter\per\second}$, the dynamic ellipse becomes a circle with $r=\SI{0.2}{\meter}$.
Which means that in high density situations the agents tend to have a circular shape instead.

A finding in \cite{Zuriguel2005jamming} is that the probability of clogs stopping the flow decreases with an increasing ratio between the size of the orifice and the size of the beads.
Therefore, we plot figure~\ref{fig:12b} with $w/r$ (the ratio between the width of the exit and the radius of the agents) as the horizontal axis.
It seems that the number of prolonged clogs is not affected by the absolute values of $w$ and $r$, provided that $w/r$ remains the same.

\section{Conclusion}
\label{sec:conclusion}
In the present paper, we focus on prolonged clogs that occur in bottleneck scenarios with the GCVM.
A general definition of prolonged clogs has been given.
Then a series of simulations in a bottleneck scenario were implemented to analyze the effect of various factors on the occurrence of prolonged clogs.

From the simulation results, the following conclusions can be drawn.
First, the number of prolonged clogs decreases as the width of the exit increases. 
Second, the occurrence of prolonged clogs cannot be eliminated by adopting a smaller time step size or updating the positions of agents sequentially.
Third, a decrease in $T$ in the GCVM leads to smaller distance between agents, which corresponds to a behavior with a higher level of motivation.
Meanwhile, decreasing $T$ reduces the space available for agents to resolve clogs, which increases the number of prolonged clogs. 
This is in accordance with the fact that clogging is more likely to occur in scenarios with a higher level of motivation.
Fourth, reducing the degree of freedom in the possible directions in which agents will move can reduce or even eliminate the occurrence of prolonged clogs.
For instance, this can be facilitated by the queuing behavior in figure.~\ref{fig:11a} as well as by locating the exit adjacent to the lower horizontal wall of the moving area.
Finally, when the ratio between the width of the exit and the radius of agents increases, the number of prolonged clogs decreases. 

\section*{Acknowledgments}
The authors are grateful to the HGF Knowledge-transfer project under Grant No. WT-0105.
Qiancheng Xu thanks the funding support from the China Scholarship Council (Grant NO. 201706060186).

\bibliographystyle{elsarticle-num}
\bibliography{main}
\end{document}